\def\grd{^{\protect\raisebox{-5pt}{$\circ$}}_{\protect\raisebox{3pt}{\,.}}}
\def\cross{\bds\times}

\def\cross{\bds\times}

\def\nk{n_{\rm b}}

\def\Pb{P_{\rm b}}

\def\sect{Section\,}
\def\fig{Figure\,}
\def\figs#1#2{Figs.\,(\ref{#1})\,to\,(\ref{#2})}
\def\rfr#1{Equation\,(\ref{#1})}
\def\rfrs#1#2{Eqs.\,(\ref{#1})\,to\,(\ref{#2})}
\def\Rfr#1{Equation\,(\ref{#1})}

\def\dert#1#2{\frac{{{\textrm{d}}}{#1}}{{{\textrm{d}}}{#2}}}

\def\virg#1{``#1"}

\def\eqi{\begin{equation}}
\def\eqf{\end{equation}}
\def\eqia{\begin{eqnarray}}
\def\eqfa{\end{eqnarray}}

\def\rp#1#2{{#1\over#2}}
\def\lb#1{\label{#1}}

\def\bds#1{\boldsymbol{#1}}

\def\ton#1{\left(#1\right)}
\def\qua#1{\left[#1\right]}
\def\grf#1{\left\{#1\right\}}

\documentclass[onecolumn]{aastex}

\usepackage{morefloats}
\usepackage[title]{appendix}
\usepackage{hyperref,textcomp}
\usepackage{booktabs}
\usepackage[table,xcdraw]{xcolor}
\usepackage{multirow}
\usepackage{rotating,tabularx}
\usepackage{float}
\usepackage{enumerate}
\usepackage{rotating}
\usepackage[polutonikogreek,english]{babel}
\usepackage{amsmath,starfont,textgreek,w-greek,wasysym}
\usepackage[flushleft]{threeparttable}
\usepackage{amsthm}
\usepackage{amscd,lineno}
\usepackage{amssymb,dsfont}
\usepackage{graphicx,epsfig}
\usepackage{txfonts}
\usepackage{xr-hyper}

\RequirePackage{color}

\newcommand{\emaila}{lorenzo.iorio@libero.it}

\allowdisplaybreaks[1]

\begin{document}

\title{Post-Keplerian obliquity variations and the habitability of rocky planets orbiting fast spinning, oblate late M dwarfs}

\shortauthors{L. Iorio}

\author{Lorenzo Iorio\altaffilmark{1} }
\affil{Ministero dell'Istruzione, dell'Universit\`{a} e della Ricerca
(M.I.U.R.)
\\ Viale Unit\`{a} di Italia 68, I-70125, Bari (BA),
Italy}

\email{\emaila}

\begin{abstract}
A couple of dozen Earth-like planets orbiting  M dwarfs  have been discovered so far. Some of them have attracted interest because of their potential long-term habitability; such a possibility is currently vigorously debated in the literature.
I show that post-Keplerian (pK) orbit precessions  may impact the habitability of a fictitious telluric planet orbiting an oblate late-type M dwarf of spectral class M9V with $M_\star=0.08\,M_\odot$ at $a=0.02\,\mathrm{au}$, corresponding to an orbital period $P_\mathrm{b}\simeq  4\,\mathrm{d}$, inducing long-term variations of the planetary obliquity $\varepsilon$  which, under certain circumstances, may not be deemed as negligible from the point of view of life's sustainability. I resume the analytical orbit-averaged equations of the pK precessions, both classical and general relativistic, of the unit vectors $\boldsymbol{\hat{S}},\,\boldsymbol{\hat{h}}$ of both the planet's spin and orbital angular momenta $\boldsymbol S,\,\boldsymbol{L}$ entering $\varepsilon$, and numerically integrate them by producing time series of the pK changes $\Delta\varepsilon\ton{t}$ of the obliquity. For rapidly rotating M dwarfs with rotational periods of the order of $P_\star \simeq 0.1-1\,\mathrm{d}$, the planet's obliquity $\varepsilon$ can undergo classical pK large variations $\Delta\varepsilon\ton{t}$ up to tens of degrees over timescales $\Delta t \simeq 20-200\,\mathrm{kyr}$, depending on the mutual orientations of the star's spin ${\boldsymbol J}_\star$, of $\boldsymbol S$, and of  $\boldsymbol L$. Instead, $\Delta\varepsilon\ton{t}$ are $\lesssim 1-1.5^\circ$ for the planet b of the Teegarden's Star. The general relativistic shifts of $\varepsilon$ are negligible. In certain circumstances, the M dwarf's oblateness $J_2^\star$ should be considered as one of the key dynamical features to be taken into account in compiling budgets of the long-term habitability of rocky planets around fast spinning late M dwarfs. My approach can be extended also to other astronomical scenarios where features of the target bodies other than their habitability are of interest.
\end{abstract}



\section{Introduction}
Since the discovery of the first\footnote{In fact, the first ever exoplanets to have been discovered were found a few years earlier around the millisecond pulsar PSR1257+12 \citep{1992Natur.355..145W}.} exoplanet around a Main Sequence (MS) star more than two decades ago \citep{1995Natur.378..355M}, the hunt for extrasolar planets\footnote{See, e.g., http://exoplanet.eu/ on the Internet.} \citep{2010exop.book.....S,2012NewAR..56....1L,2018haex.bookE....D,2018exha.book.....P} and their physical characterization has become an increasingly active effort, especially in view of the possibility that some of them may be habitable \citep{2013Sci...340..577S,2017ARA&A..55..433K,Meadows2018,2018NatAs...2..432S,2018AsBio..18..663S,2019RvMP...91b1002L,2019Sci...364..434S,2020Univ....6..130I}.
As a consequence, the novel multidisciplinary field of research known as astrobiology \citep{2020IJAsB..19..379L} has flourished.

The purpose of this paper is to illustrate that, under certain environmental circumstances and in a specific \textcolor{black}{Newtonian} dynamical way, \textcolor{black}{the quadrupole mass moment $J_2^\star$ of the M dwarf,} should be counted among the several effects of different nature concurring to constrain the capability of a certain class of exoplanets of hosting and sustaining life and, possibly, civilizations over timescales of geological length. \textcolor{black}{Also the impact of the Einstein's General Theory of Relativity (GTR) \citep{2016Univ....2...23D,2017grav.book.....M} will be investigated, despite it will be found to be negligible, at least in this specific astronomical scenario.}

To be more specific, I will focus on planets analogous to those, listed in the online Habitable Exoplanets Catalogue maintained  at http://phl.upr.edu/projects/habitable-exoplanets-catalog by the University of Puerto Rico at Arecibo, that are more likely to have a rocky composition and maintain surface liquid water, characterized by masses $M$ and radii $R$ in the range\footnote{The labels $\oplus$ and $\odot$ appended to $M,\,R$ refer to Earth and  Sun, respectively.} $0.1\,M_\oplus\lesssim M\lesssim 5\,M_\oplus$ and $0.5\,R_\oplus\lesssim R\lesssim 1.5\,R_\oplus$, typically orbiting at $r\simeq 0.02\,\mathrm{astronomical\,units}\,\ton{\mathrm{au}}$ from M dwarfs having masses $M_\star$ and radii $R_\star$ of the order of $M_\star\simeq 0.08\,M_\odot,\,R_\star\simeq 0.1\,R_\odot$. The habitability of such a class of exoplanets has been the subject of intense scrutiny so far \citep{1964hpfm.book.....D,2007AsBio...7...30T,1999OLEB...29..405H,2016PhR...663....1S,2020ApJ...888..102L}. It may be useful to recall that M dwarfs are MS red dwarf stars of spectral class M whose mass ranges from $0.6\,M_\odot$ down to about $0.08\,M_\odot$ \citep{2000ApJ...542..464C}, which seems to be the minimum value of the stellar mass allowing to sustain stable hydrogen burning. M dwarfs are the most numerous stars in the Galaxy, contributing to approximately $40\%$ of its total stellar mass \citep{1996ApJ...465..759G,1996ApJ...459L..87M}. The stars with $0.075\,M_\odot\lesssim M_\star \lesssim 0.1\,M_\odot$ are usually dubbed as late M dwarfs (\textcolor{black}{M9V}-M6V spectral class), and are fully convective \citep{1991AJ....101..662B}. The MS lifetime of late-type M dwarfs with $M_\star = 0.08\,M_\odot$ can be as long as $\simeq 12\,\mathrm{Gyr}$ \citep{2005AN....326..913A}.

The general relativistic effects I refer to are the de Sitter-Fokker \citep{1916MNRAS..77..155D,1918KNAB...27.214S,1921KNAB...23..729F}, or geodetic, and the Pugh-Schiff \citep{Pugh59,Schiff60} precessions of the spin of a pointlike gyroscope freely orbiting a localized mass-energy source such as an astronomical spinning body. To the post-Newtonian\footnote{The pN expansion is one of the most successful and widely used approximation scheme to solve the fully nonlinear Einstein's equations to describe motions of arbitrary shaped, massive bodies \citep{1997PThPS.128..123A,2003Blanchet,2018tegp.book.....W}.} (pN) level, the geodetic precession is due to the  static \virg{gravitoelectric} deformation of the spacetime induced by the mass monopole moment, i.e. the mass $M$, of the source, assumed nonrotating. Instead, the Pugh-Schiff spin rate is caused by the stationary contribution  of the source's spin dipole moment, i.e. its proper angular momentum $\bds J$, and is often dubbed as \virg{gravitomagnetic}. Such appellations have nothing to do with the electric and magnetic fields generated by electric charges and currents, being  due to the purely formal resemblance of the linearized  Einsteinian field equations, valid in the slow-motion and weak-field limit, with the equations of the Maxwellian electromagnetism \citep{Thorne86,2001rfg..conf..121M,2001rsgc.book.....R}. The geodetic spin precession has been successfully measured in a number of scenarios ranging from the dedicated space-based Gravity Probe B (GP-B) mission in the field of Earth \citep{2011PhRvL.106v1101E,2015CQGra..32v4001E} and the Earth-Moon motion around the Sun \citep{1996PhRvD..53.6730W,2009IAU...261.0801W,2018CQGra..35c5015H}, monitored with the Lunar Laser Ranging (LLR) technique \citep{1994Sci...265..482D}, to some binary pulsars \citep{2008Sci...321..104B,Kramer2012}; the most accurate tests so far, accurate to the $\simeq 10^{-3}$ level, are the GP-B and LLR ones. The much smaller gravitomagnetic Pugh-Schiff effect has been measured so far only by GP-B with a $\simeq 19\%$ accuracy \citep{2011PhRvL.106v1101E,2015CQGra..32v4001E}; \citet{1975Ap&SS..32....3H} suggested to measure the spin angular momenta $\bds J$ of the Sun and Jupiter by exploiting the Pugh-Schiff effect with dedicated spacecraft-based missions, but such a proposal was never carried out. For a recent recap of the de Sitter and Pugh-Schiff spin precessions, see, e.g., \citet{2014grav.book.....P}.

In the case of the aforementioned extrasolar scenario, the \textcolor{black}{post-Keplerian (pK)} precessions of the exoplanet's spin ${\bds S}$ due to the \textcolor{black}{non-spherically symmetric classical quadrupolar} field of its close parent star\textcolor{black}{, along with its pN gravitoelectromagnetic components,} may affect, in principle, its obliquity $\varepsilon$, which is defined as the angle between ${\bds S}$ and the orbital angular momentum $\bds L$ or, equivalently, between the planet's equator and its orbital plane. It is important to remark that the \textcolor{black}{pK} spin precessions are not the only dynamical effects to be considered in assessing potential long-term changes of $\varepsilon$. Even by limiting to a purely two-body scenario involving only the host star and a planet, the axisymmetric pK components  of  the stellar  gravitational field, both of Newtonian and pN nature, let the orbital plane undergo long-term variations. Indeed, $J_2^\star$  induces classical long-terms rates of change of \textcolor{black}{$\bds L$} \citep{1975PhRvD..12..329B}. Moreover, to the pN level, also the gravitomagnetic Lense-Thirring effect \citep{1918PhyZ...19..156L} due to ${\bds J}_\star$ further contributes to displacing \textcolor{black}{$\bds{L}$} \citep{1975PhRvD..12..329B}. Depending on the magnitude of the resulting long-term shift $\Delta\varepsilon\ton{t}$,  the habitability of the planet may be more or less affected since the obliquity is a key factor in constraining it over the \ae ons. Indeed, it controls the insolation received from the parent star \citep{1993Natur.361..615L,1997Icar..129..254W,2004A&A...428..261L,2014AsBio..14..277A,2015P&SS..105...43L,
2017arXiv171008052Q,2017ApJ...844..147K,2018AJ....155..237S,2019arXiv191108431Q}. The obliquity of Earth $\varepsilon_\oplus$  changes slowly with time from $\simeq 22\grd1$ to $24\grd5$, undergoing an oscillation cycle with amplitude $\lesssim 2\grd4$ in about $41\,\mathrm{kyr}$ \citep{2019arXiv191108431Q}. The value of $\varepsilon_\oplus$  impacts the seasonal cycles and its long-term variation affects directly the terrestrial climate \citep{Milan1941}, as deduced from geologic records \citep{Kerr1987,1995GeoJI.121...21M,1999E&PSL.174..155P}.  An interesting feature is that large planetary obliquity  increases planet's seasonality  \citep{1974JGR....79.3375W,1975GeoM..112..441W,2003IJAsB...2....1W,2013Icar..226..760D,2016PhR...663....1S}. Furthermore, it has been recently shown that habitable surface area increases for large obliquities \citep{2009ApJ...691..596S,2016PhR...663....1S}. As far as Earth-like planets orbiting M-dwarfs are concerned, it was shown that their obliquity may play a role in regulating atmospheric escape \citep{2019ApJ...882L..16D}. Long-term variations of the obliquity $\Delta\varepsilon(t)$ can drive important modifications in planetary climate; currently, a debate exists about the negativity or otherwise of their impact on habitability. If, on the one hand, rapid and/or large changes $\Delta\varepsilon(t)$ may induce climate shifts which can be commensurately severe, as remarked, e.g., by \citet{2004Icar..171..255A}, on the other hand, exoplanets experiencing high-frequency oscillations in obliquity may avoid global glaciation, since neither pole of the planet faces away from the star for a long enough time for thick ice sheets to develop \citep{2014AsBio..14..277A,2016PhR...663....1S}. Tidally-locked M-dwarf planets with non-zero obliquities could increase surface temperatures because of reduced negative cloud feedback \citep{2016ApJ...823L..20W,2016PhR...663....1S}. If other major bodies are present as well, \textcolor{black}{both the spin and} the orbital plane of the planet \textcolor{black}{are}, in general, displaced \textcolor{black}{over long time spans} also by their gravitational pull \textcolor{black}{\citep[e.g.][]{2011CeMDA.111..105C}}. Moreover, the presence of massive moons may affect the planetary obliquity in a way such that its impact on habitability is still debated \citep{2012Icar..217...77L,2014ApJ...790...69L,2014IJAsB..13..324S,2016PhR...663....1S}. Here, I will restrict to a restricted two-body scenario. It is worthwhile noticing that the \textcolor{black}{pK} variations of obliquity treated here are independent of the physical features of the planet itself and on the presence of moon or other planets, depending only on the nature and the size of the deformation of the spacetime around the host star. \textcolor{black}{Nonetheless, even by neglecting the possible presence of further bodies, other dynamical features depending on the specific physical properties of the planet itself may, in principle, have long-term effects on $\varepsilon$. Indeed, to the Newtonian level, the oblateness $J_2$ of the planet, both of centrifugal and tidal origin, affects both $\bds S$ and $\bds L$ \citep{1975PhRvD..12..329B}. Moreover, also tidal effects,  parameterized in terms of the planet's Love number $k_2$ and the time lag $\Delta t$ caused by the dissipation of mechanical energy of tides in the planet's interior \citep{1964RvGSP...2..661K}, induce, among other things, long-term variations on the obliquity \citep[e.g.][]{2011CeMDA.111..105C}.
The treatment of such planet-dependent, potentially relevant features is beyond the scopes of the present paper.
}

I will show that, in the mutual interplay of the pK orbital and spin precessions, a key parameter turns out to be the rotational period\footnote{The minimum breakup period for the M dwarf considered here is $P_\star^\mathrm{break}=2\,\uppi\,\sqrt{R^3_\star/\mu_\star}=0.01\,\mathrm{d}$. } $P_\star$ of the star
entering the calculation of its dimensionless quadrupole mass moment $J_2^\star$ and spin angular momentum ${\bds J}_\star$.
Indeed, it turns out that, for $0.1\,\mathrm{d}\lesssim P_\star\lesssim 1\,\mathrm{d}$,
the resulting \textcolor{black}{$J_2^\star$-driven pK} variations $\Delta\varepsilon\ton{t}$ of the planet's obliquity may be as large as tens of degrees with characteristic timescales less than $1\,\mathrm{Myr}$ or so. \textcolor{black}{Instead, the magnitude of the pN changes of the obliquity, characterized by longer characteristic time scales, is too small to have an impact.}

In order to properly put into context my findings and assess their significance, it is legitimate to wonder if the fast rotation of M dwarfs lasts  enough to be significative for the emergence of life and, possibly, civilizations on planets orbiting them, and if short rotational periods of such relatively dim stars are somehow connected with some physical mechanisms which may be harmful for life on their rocky companions. Moreover, despite the goal of my work is \textcolor{black}{neither} performing a test of GTR \textcolor{black}{nor measuring or constraining $J_2^\star$ by means of}  some existing natural probes, it may also be interesting to know if planets orbiting fast rotating M dwarfs are, actually, detectable. To these aims, the following considerations are in order.

The rotational period $P_\star$ of M dwarfs can be observationally determined in two ways \citep{2015ApJ...812....3W}: by measuring the periodicity of brightness variations induced by the presence of photospheric spots coming in to and out of view as the star rotates (photometric rotation period) \citep{2007AcA....57..149K}, and  by looking for velocity broadening of spectral lines through the projected rotation velocity\footnote{The angle $i_\star$ is the inclination of the star's spin  ${\bds J}_\star$ to the line of sight such that $i_\star=90\circ$ corresponds to an edge-on view, while $i_\star=0^\circ$ means that the star's equatorial plane is perpendicular to the line of sight.} $u_\star\doteq v_\star\sin i_\star$. In fact, several photometric periods as short as $P_\star\simeq 0.1-1\,\mathrm{d}$ have been measured so far for many M-dwarfs \citep{2014ApJ...788..114R,2016ApJ...821...93N,2018AJ....156..217N,2020MNRAS.491.5216G}. As far as spectroscopical determinations are concerned, rotational periods of the order of $P_\star \simeq 0.1-1\,\mathrm{d}$ correspond to $u_\star\simeq 4.5-30\,\mathrm{km\,s}^{-1}$ for, say, $i_\star\simeq 50-60^\circ$ and $R_\star = 0.1\,R_\odot$; similar values of $u_\star$ were, in fact, measured for a number of M dwarfs \citep{2012ApJS..203...10T,2016ApJ...821...93N,2018A&A...612A..49R}, \textcolor{black}{like, e.g., LP 944-20, a late-M dwarf of spectral class M9V, having $u_\star=30.8\,\mathrm{km\,s}^{-1}$ \citep{2015ApJ...812...42B}. }. The time short rotation periods persist depends on mass, and is likely close to $\simeq 5\,\mathrm{Gyr}$ for late-type M dwarfs\footnote{E.R. Newton, personal communication. \textcolor{black}{Except in the cases of binaries, rotation declines with age; however, the timeline for how this occurs in late-M dwarfs is not clear.}}.
\textcolor{black}{This seems likely to be the case especially for M9V stars with $M_\star\simeq 0.075-0.08\,M_\odot$. It is no longer true for slightly more massive M6V dwarfs with $M_\star=0.1\,M_\odot$ whose magnetic breaking leads to large angular momentum losses over time, so that $P_\star\simeq 40\,\mathrm{d}$ after 2 Gyr \citep{2018RNAAS...2...34E}. Mid-to-late M dwarfs reach periods of 100 d or more by a typical age of 5 Gyr \citep{2016PhR...663....1S}. } The earliest fossil evidence suggests that it took $\simeq 0.8-1\,\mathrm{Gyr}$ for life to develop
on our planet \citep{Aw92,1996Natur.384...55M,2007Geo....35..591M,2016Natur.537..535N,2016PhR...663....1S}, or even less ($0.5\,\mathrm{Gyr}$) \citep{Bell14518,2016PhR...663....1S}.

On the other hand, fast rotation and stronger magnetic activity inducing, e.g., X-ray and UV (XUV) flares are linked \citep{2016A&A...595A..12S}: fast rotating M dwarfs are the most likely to flare \citep{2020AJ....159...60G}. Nonetheless, such a correlation, though still present, is less pronounced in fully convective, late-type M dwarfs \citep{2015ApJ...812....3W,2016PhR...663....1S}.  If, on the one hand, XUV flares may be detrimental to life on otherwise potentially habitable planets because of atmospheric erosion, destruction of the ozone layers on oxic planets,
stress for surface life, on the other hand, they could trigger the genesis of life  powering prebiotic chemistry, producing surface biosignatures, or serving as a source for otherwise scarce visible-light photosynthesis on planets orbiting M dwarfs; see the discussion in, e.g., \citet[Section 1]{2020AJ....159...60G}, and references therein.

As far as the detectability of planets around fast spinning M dwarfs is concerned, it may be challenging for the radial velocity method \citep{2018haex.bookE..16B,2018haex.bookE..18M,2018haex.bookE...4W} because of the broadening of the spectral lines. On the other hand, the transit method \citep{2009ApJ...705..683B} can still readily detect planets around rapidly rotating stars \citep{2018haex.bookE.117D}.

Conversely, the same calculation, applied to the existing exoplanet Teegarden's Star b \citep{2019A&A...627A..49Z} shows that its \textcolor{black}{pK} obliquity changes $\Delta\varepsilon\ton{t}$ are likely too small to have any noticeable impact on its habitability.

The  paper is organized as follows. In \sect\ref{sec.2}, I \textcolor{black}{review} the \textcolor{black}{analytical} orbit-averaged equations for the \textcolor{black}{Newtonian and pN} time variations of $\bds{\hat{S}}$ and of the orbital plane required to calculate the change $\Delta\varepsilon$ of the planetary obliquity. \sect\ref{sec.3} is devoted to the calculation of the centrifugal quadrupole mass moment $J_2^\star$ and of the spin angular momentum $J_\star$ of the M dwarf entering the aforementioned equations. My numerical integrations of the coupled equations for the \textcolor{black}{pK} time variations of $\bds{\hat{S}}$ and of the orbital plane, and the resulting time series for the variations $\Delta\varepsilon\ton{t}$  are displayed in \sect\ref{sec.4}\textcolor{black}{; it displays both purely pN and classical runs in order to disentangle what comes just from $J_2^\star$ and what comes from GTR}. \sect\ref{sec.5} summarizes my findings and offers my conclusions.
\section{\textcolor{black}{The mathematical model for the evolution of the planetary spin's obliquity}}\lb{sec.2}
\textcolor{black}{
The time-dependent evolution of the obliquity $\varepsilon\ton{t}$ can be obtained from, e.g.,
\eqi
\cos\varepsilon\ton{t} = \bds{\hat{S}}\ton{t}\bds\cdot\bds{\hat{h}}\ton{t}.\lb{oblateness}
\eqf
Thus, one needs to know how both $\bds{\hat{S}}$ and $\bds{\hat{h}}$ evolve driven by the quadrupolar classical field of the star and GTR.
}

\textcolor{black}{The total pN rate of change of $\bds{\hat{S}}$, averaged over one orbital period $\Pb$ of the planet about its parent star,
\eqi
\dert{\bds{\hat{S}}}{t} = {\mathbf{\Omega}}_\mathrm{pN}\bds\cross\bds{\hat{S}},\lb{dSdtpN}
\eqf
can be inferred from the sum of the averaged gravitoelectric (GE) and gravitomagnetic (GM) spin precessions
\eqi
{\mathbf{\Omega}}_\mathrm{pN} \doteq {\mathbf{\Omega}}_\mathrm{GE} + {\mathbf{\Omega}}_\mathrm{GM},
\eqf
where, in the test particle limit\textcolor{black}{\footnote{\textcolor{black}{Here and in the following, the star is assumed to be the body $\ton{2}$ of \citet{1975PhRvD..12..329B}, while the planet is the body $\ton{1}$, so that the approximation $m_2\gg m_1$ pertaining their masses can be assumed in all the relevant formulas by \citet{1975PhRvD..12..329B}.}}}, \citep{1975PhRvD..12..329B}
\begin{align}
{\mathbf{\Omega}}_\mathrm{GE} \lb{dst} & = \rp{3\,\nk\,\mu_\star}{2\,c^2\,a\,\ton{1-e^2}}\bds{\hat{h}}, \\ \nonumber \\
{\mathbf{\Omega}}_\mathrm{GM} \lb{ltr} & =\rp{G\,J_\star}{2\,c^2\,a^3\,\ton{1-e^2}^{3/2}}\qua{{\bds{\hat{J}}}_\star - 3\,\ton{{\bds{ \hat{J}}}_\star\bds\cdot\bds{\hat{h}}}\,\bds{\hat{h}}}.
\end{align}
In \rfrs{dst}{ltr}, $c$ is the speed of light in vacuum, $G$ is the Newtonian gravitational constant, $\mu_\star\doteq GM_\star$ is the star's gravitational parameter, $a,\,e$ are the semimajor axis and the eccentricity, respectively, of the planet's astrocentric orbit, $\nk=\sqrt{\mu_\star/a^3}$ is the Keplerian mean motion,  $\bds{\hat{h}}$ is the unit vector, perpendicular to the plane of the planetary orbital motion, directed along the orbital angular momentum $\bds L$.}

\textcolor{black}{In principle, the spin axis ${\bds{\hat{J}}}_\star$ of an oblate star may experience a long-term, orbit-averaged classical precession induced by the planet \citep[e.\,g.][]{2011CeMDA.111..105C}; it can be expressed as \citep{1975PhRvD..12..329B}
\eqi
\dert{{\bds{\hat{J}}}_\star}{t} = \mathbf{\Omega}_\mathrm{N}\bds\times{\bds{\hat{J}}}_\star,\lb{dJdt}
\eqf
with\textcolor{black}{\footnote{\textcolor{black}{The dimensional quantity $\Delta I^{\ton{1,\,2}}$ for the body $\ton{1}$ or $\ton{2}$ entering \citet[Eq.\,$\ton{47}$]{1975PhRvD..12..329B} corresponds to $J_2^\star\,M_\star\,R_\star^2$ in the case of the star.}}} \citep{1975PhRvD..12..329B}
\eqi
\mathbf{\Omega}_\mathrm{N}=\rp{\mu_\star\,M\,J_2^\star\,R^2_\star}{2\,J_\star\,a^3\,\ton{1-e^2}^{3/2}}\qua{{\bds{\hat{J}}}_\star - 3\,\ton{{\bds{ \hat{J}}}_\star\bds\cdot\bds{\hat{h}}}\,\bds{\hat{h}}}.\lb{booh}
\eqf
Nonetheless, in the present scenario, the planet-driven shift of ${\bds{\hat{J}}}_\star$ turns out to be negligible.
}

\textcolor{black}{As far as $\bds{\hat{h}}$ is concerned, its pN evolution is driven by
\eqi
\dert{\bds{\hat{h}}}{t} = {\mathbf{\Psi}}_\mathrm{pN}\bds\cross\bds{\hat{h}},\lb{dLdtpN}
\eqf
with
\eqi
{\mathbf{\Psi}}_\mathrm{pN} \doteq {\mathbf{\Psi}}_\mathrm{GE} + {\mathbf{\Psi}}_\mathrm{GM},
\eqf
where, in the test particle limit, \citep{1975PhRvD..12..329B}
\begin{align}
{\mathbf{\Psi}}_\mathrm{GE} \lb{dstL} & = 2\,{\mathbf{\Omega}}_\mathrm{GE}, \\ \nonumber \\
{\mathbf{\Psi}}_\mathrm{GM} \lb{ltrL} & =4\,\,{\mathbf{\Omega}}_\mathrm{GM}.
\end{align}
In fact, $\bds{\hat{h}}$ is changed also to the Newtonian order by other dynamical features like the stellar quadrupole moment $J_2^\star$ \citep[e.\,g.][]{2011CeMDA.111..105C} according to
\eqi
\dert{\bds{\hat{h}}}{t} = {\mathbf{\Psi}}_\mathrm{N}\bds\cross\bds{\hat{h}},\lb{dLdtJ2}
\eqf
where, in the test particle limit\textcolor{black}{\footnote{\textcolor{black}{In the notation of \citet[eq.\,(72)]{1975PhRvD..12..329B}, $J_2$ is dimensionally the square of a length corresponding, actually,  to $J_2\,R^2$.}}}, \citep{1975PhRvD..12..329B}
\eqi
{\mathbf{\Psi}}_\mathrm{N}= -\rp{3\,\nk\,J_2^\star\,R^2_\star}{4\,a^2\,\ton{1-e^2}^2}\,\grf{2\ton{{\bds{\hat{J}}}_\star\bds\cdot\bds{\hat{h}}}\,{\bds{\hat{J}}}_\star + \qua{1 - 5\ton{{\bds{\hat{J}}}_\star\bds\cdot\bds{\hat{h}}}^2}\,\bds{\hat{h}}  }.\lb{psiN}
\eqf
}

\textcolor{black}{The sum of \rfr{dLdtpN} and \rfr{dLdtJ2}, which can be posed as
\eqi
\dert{\bds{\hat{h}}}{t} = \mathbf{\Psi}\bds\times\bds{\hat{h}},\lb{dLdt}
\eqf
with
\eqi
\mathbf{\Psi}\doteq \mathbf{\Psi}_\mathrm{N}+\mathbf{\Psi}_\mathrm{pN},
\eqf
and \rfr{dSdtpN} can be simultaneously integrated in order to produce time series of $\varepsilon\ton{t}$ according to \rfr{oblateness}. To this aim, the initial conditions for $\bds{\hat{S}}$ and $\bds{\hat{h}}$ can be conveniently parameterized as
\begin{align}
{\hat{S}}_x \lb{Sx} & = \sin\theta_0\,\cos\alpha_0, \\ \nonumber \\
{\hat{S}}_y & = \sin\theta_0\,\sin\alpha_0, \\ \nonumber \\
{\hat{S}}_z \lb{Sz}& = \cos\theta_0, \\ \nonumber \\
{\hat{h}}_x \lb{Lx} & = \sin I_0\,\sin\Omega_0, \\ \nonumber \\
{\hat{h}}_y & = -\sin I_0,\,\cos\Omega_0 \\ \nonumber \\
{\hat{h}}_z \lb{Lz}& = \cos I_0,
\end{align}
so that $\alpha_0$ is the planetary spin's azimuthal angle  in the  $\Pi=\grf{x,\,y}$ reference plane of the adopted coordinate system, and $\theta_0$ is its colatitude: $\theta_0 = 0^\circ$ implies that $\bds{\hat{S}}$ is perpendicular to $\Pi$ at the initial epoch $t_0$. Furthermore, $\Omega_0$ is the longitude of the ascending node of the planet's orbit, and $I_0$ is the inclination of the orbital plane to $\Pi$; $I_0=0^\circ$ implies that just the ecliptic plane at epoch is chosen as reference plane, so that $\theta_0=\varepsilon_0$.
The star's spin axis, assumed as constant during the integration of \rfr{dLdt} and \rfr{dSdtpN}, can be similarly parameterized as
\begin{align}
{\hat{J}}^\star_x \lb{Jx} & = \sin\eta_\star\,\cos\varphi_\star, \\ \nonumber \\
{\hat{J}}^\star_y & = \sin\eta_\star\,\sin\varphi_\star, \\ \nonumber \\
{\hat{J}}^\star_z \lb{Jz}& = \cos\eta_\star.
\end{align}
  }

\textcolor{black}{From \rfr{dSdtpN}, \rfr{dJdt}, \rfr{dLdtJ2}, \rfr{dLdt}, calculated with  \rfrs{dst}{ltr}, \rfr{booh} and \rfr{psiN},
it can be noted that, if the three angular momenta are exactly aligned, they stay constant.
}
\section{The stellar quadrupole mass moment and angular momentum}\lb{sec.3}
Two key parameters entering \textcolor{black}{\rfr{ltr}, \rfr{ltrL}, \rfr{booh} and \rfr{psiN}}  are the dimensionless quadrupole mass  moment $J_2^\star$ and the spin angular momentum $J_\star$ of the M dwarf, for which, to my knowledge, no estimates exist in the literature. Thus, in the following, I will infer some plausible guesses for their values, to be used in the subsequent numerical integrations.

The first even zonal harmonic $J_2^\star$ of the multipolar expansion of the non-spherical component of the Newtonian gravitational potential of a MS star can be expressed as \citep{2009ApJ...698.1778R,2011A&A...528A..41L}
\eqi
J_2^\star = \rp{k^\star_2}{3}\,q_\star.\lb{J2}
\eqf
In \rfr{J2}, $k_2^\star$ is the Love number \citep{1939MNRAS..99..451S,1959cbs..book.....K,2009ApJ...698.1778R,2011A&A...528A..41L}, while the parameter $q_\star$ is defined as
\citep{2009ApJ...698.1778R,2011A&A...528A..41L}
\eqi
q_\star\doteq\rp{\omega^2_\star\,R_\star^3}{\mu_\star},\lb{qu}
\eqf
where $\omega_\star \doteq 2\uppi/P_\star$ is the star's angular speed.
Thus, one has
\eqi
J_2^\star = \rp{4\,\uppi^2\,k_2^\star\,R_\star^3}{3\,P^2_\star\,\mu_\star}.\lb{J2Pstar}
\eqf

If the spectroscopic period is available, it is convenient to derive an expression of $P_\star$ in terms of $u_\star$. By writing the stellar rotational period $P_\star$   as
\eqi
P_\star = \rp{2\uppi\,R_\star}{v_\star}=\rp{2\uppi\,R_\star\sin i_\star}{u_\star},
\eqf
the stellar angular speed $\omega_\star$ can be cast into the form
\eqi
\omega_\star =\rp{u_\star}{R_\star\sin i_\star}.
\eqf
Thus, from \rfrs{J2}{qu}, one finally has
\eqi
J_2^\star = \rp{k_2^\star\,R_\star\,u^2_\star}{3\,\mu_\star\sin^2 i_\star}.\lb{J2q}
\eqf
\rfr{J2Pstar} and \rfr{J2q} express $J_2^\star$ in terms of quantities like $M_\star,\,R_\star,\,P_\star,\,u_\star$, and, sometimes, $i_\star$ as well,  which are measured for many M dwarfs, and can be retrieved in the literature.

The stellar spin angular momentum can approximately be expressed as
\eqi
J_\star= \mathcal{J}_\star\,M_\star\,R_\star^2\,\omega_\star,\lb{Gei}
\eqf
where $\mathcal{J}_\star$ is the moment of inertia factor which, according to the Darwin-Radau equation, is given by \citep{2004A&A...428..691B}
\eqi
\mathcal{J}_\star = \rp{2}{3}\,\ton{1-\rp{2}{5}\sqrt{1+\beta_\star}},
\eqf
with
\eqi
\beta_\star \doteq \rp{5\,q_\star}{2\,f_\star}-2.\lb{beta}
\eqf
In \rfr{beta}, $f_\star$ is the geometric flattening, defined as the difference between the star's equatorial and polar radii normalized to the equatorial radius, which, in this case, is given by \citep{2011A&A...528A..41L}
\eqi
f_\star = \ton{\rp{k^\star_2 +1 }{2}}\,q_\star.
\eqf
Thus, the stellar angular momentum can be finally expressed as
\eqi
J_\star = \rp{4\,\uppi}{15}\,\ton{5-2\sqrt{\rp{5}{1+k^\star_2} -1 }}\,\rp{M_\star\,R^2_\star}{P_\star},\lb{JPstar}
\eqf
or
\eqi
J_\star = \rp{2}{15}\,\ton{5-2\sqrt{\rp{5}{1+k^\star_2} -1 }}\,\rp{M_\star\,R_\star\,u_\star}{\sin i_\star}.\lb{Jstar}
\eqf
Also \rfrs{JPstar}{Jstar} $M_\star,\,R_\star,\,P_\star,\,u_\star$, and, sometimes, $i_\star$ as well.

Concerning the Love number $k_2$ entering both $J_2^\star$ and $J_\star$, its possible values range from 0 to $1.5$, the latter figure holding for the limiting case of a sphere with homogeneous density \citep{2018AJ....156..149B}. In general, a concentration of mass toward the center results in a smaller $k_2$ value \citep{2018AJ....156..149B}.
For a Sun-type MS star, it is of the order of \citep{2009ApJ...698.1778R}
\eqi
k_2^\star\simeq 0.03.\lb{kappa2}
\eqf
For a late M dwarf with $M_\star = 0.08\,M_\odot$ and the same metallicity as the Sun, A. Claret\footnote{Personal communication, February 2021.} was able to calculate an evolutionary track  beginning at the Pre-Main Sequence (PMS) with his code developed in \citet{2004A&A...424..919C}. The values of $k_2^\star$ depend on the age, but they notably differ from \rfr{kappa2}. In the MS, it turns out to be of the order of
\eqi
k^\star_2\simeq 0.16, \lb{claret}
\eqf
corresponding to totally convective models. \textcolor{black}{Table 2 of \citet{2009A&A...494..209L} lists, among other things, $\log k_2^\star$ and the radius of gyration $\sqrt{\mathcal{J}_\star}$ for a rotating Zero Age Main Sequence (ZAMS) star with $M_\star=0.09\,M_\odot,\,P_\star\simeq 0.03\,\mathrm{d}$ getting $k_2^\star\simeq 0.15$ and $\mathcal{J}_\star\simeq 0.216$.}
It may be interesting a comparison \textcolor{black}{also} with some brown dwarfs.
For\footnote{The masses of the brown dwarfs Gliese-229b and range within $0.062-0.071\,M_\odot$ \citep{2020AJ....160..196B} and $0.02\,M_\odot$ \citep{2008A&A...491..889D}, respectively, while the mass of the M-dwarf Teegarden's Star is $M_\star=0.089\,M_\odot$.} Gliese-229b \citep{1995Natur.378..463N} and Corot-3b \citep{2008A&A...491..889D},  \citet{2018AJ....156..149B} calculated $k_2 = 0.349$ and $k_2 = 0.387$, respectively, which are almost twice the value of \rfr{claret}.

As far as the projected rotational velocity $u_\star$ is concerned, it turns out that $u_\star\simeq 2-20\,\mathrm{km\,s}^{-1}$ \citep{2012ApJS..203...10T,2018A&A...612A..49R} for several M-dwarfs.

\begin{figure*}[ht]
\centering
\begin{tabular}{cc}
\includegraphics[width=8cm]{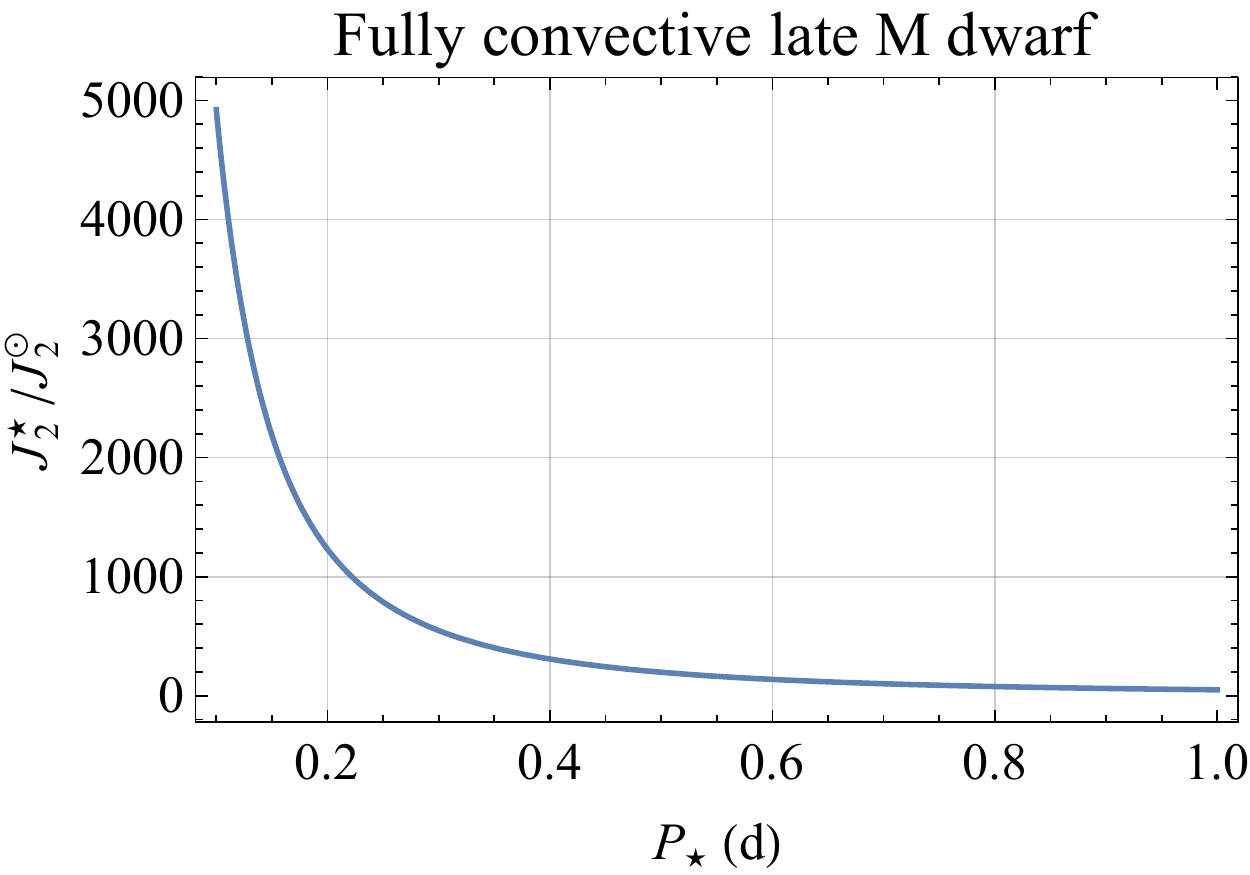}&\includegraphics[width=8cm]{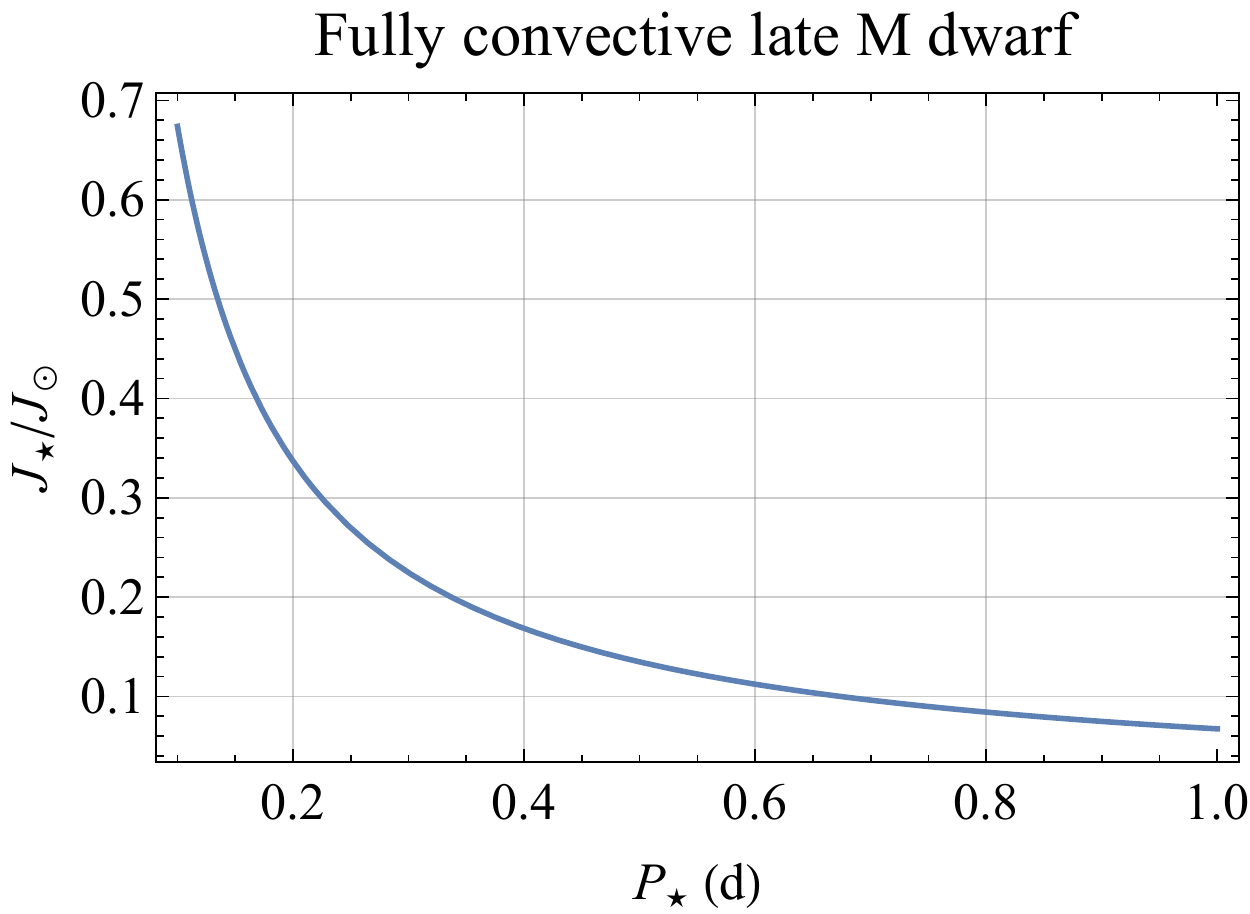}\\
\end{tabular}
\caption{
Plot of $J_2^\star$ and $J_\star$, normalized to the corresponding values of the Sun $J_2^\odot = 2.2\times 10^{-7}$ \citep{2018NatCo...9..289G} and $J_\odot = 1.92\times 10^{41}\,\mathrm{J\,s}$ \citep{1998MNRAS.297L..76P}, of a fully convective late M dwarf with $M_\star=0.089\,M_\odot,\,R_\star=0.107\,R_\odot,\,k_2^\star=0.16$ as a function of the star's rotational period $P_\star$, in d, according to \rfr{J2Pstar} and \rfr{JPstar}.
}\label{figura0}
\end{figure*}

In \fig\ref{figura0}, I plot $J_2^\star,\,J_\star$, normalized to the corresponding values of the Sun, for a fully convective late M dwarf with $M_\star=0.089\,M_\odot,\,R_\star=0.107\,R_\odot,\,k_2^\star=0.16$  as a function of the star's rotational period $P_\star$.
\section{Numerical calculation}
\begin{figure*}[ht]
\centering
\begin{tabular}{cc}
\includegraphics[width=8cm]{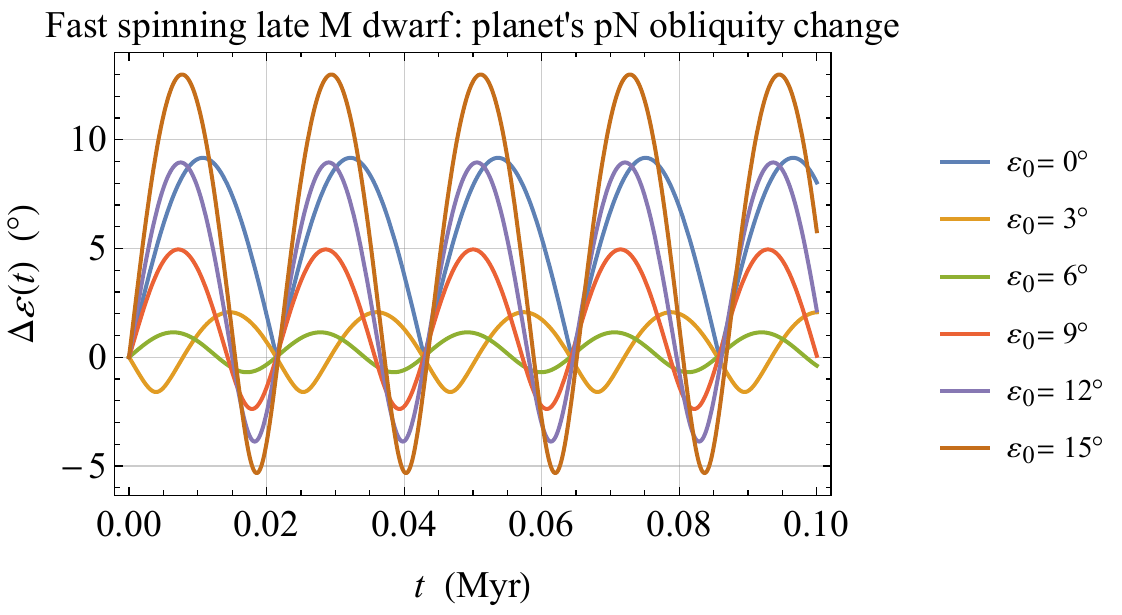} & \includegraphics[width=8cm]{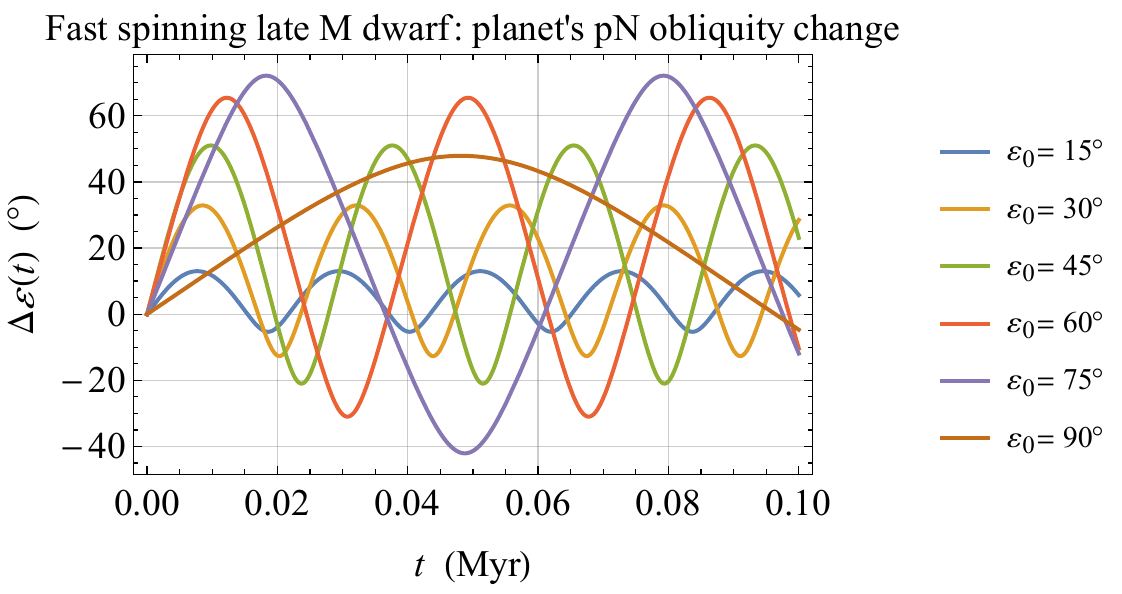}\\
\includegraphics[width=8cm]{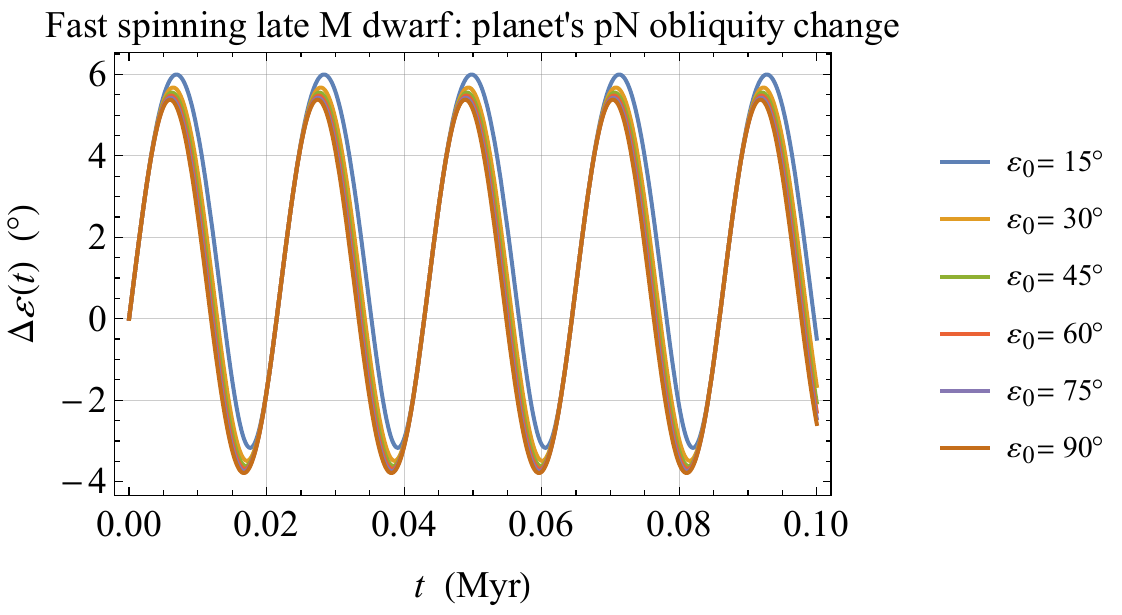} & \includegraphics[width=8cm]{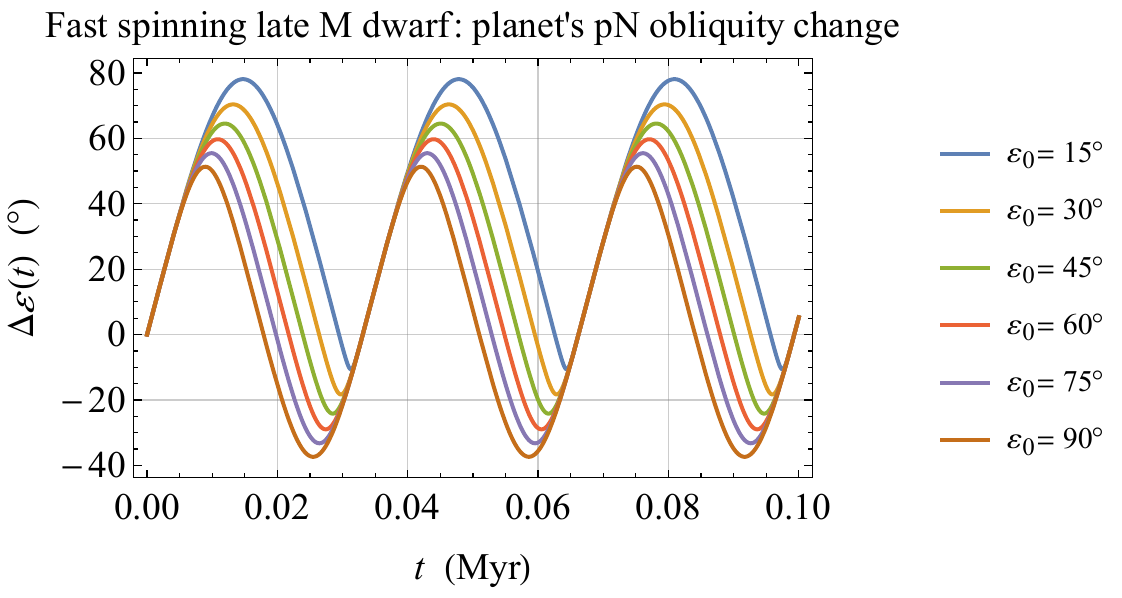}\\
\end{tabular}
\caption{
Numerically integrated time series of the \textcolor{black}{$J_2^\star$-driven pK} variation $\Delta\varepsilon(t)$ of the time-dependent obliquity $\varepsilon(t)$ of the spin axis $\bds{\hat{S}}$ of a fictitious potentially habitable planet orbiting a fast spinning late-type M dwarf characterized by $M_\star=0.089\,M_\odot,\,R_\star=0.107\,R_\odot,\,k_2^\star=0.16,\,P_\star = 0.1\,\mathrm{d}$ and different initial orientations of the planetary and stellar spin angular momenta $\bds{\hat{S}},\,\bds{\hat{J}}_\star$, all referred to the ecliptic plane at the initial epoch as reference $\grf{x,\,y}$ plane. \Rfr{oblateness} was used after integrating \textcolor{black}{\rfr{dLdtJ2}}. The initial values adopted for the relevant spin and orbital parameters, common to all the integrations, are $e=0.0,\,a=0.0252\,\mathrm{au},\,I_0=\textcolor{black}{0.0^\circ},\,\Omega_0=50^\circ,\,\alpha_0=50^\circ,\,\varphi_\star=150^\circ$.  In the upper left panel, moderate spin-orbit misalignments between the initial configurations of ${\bds{\hat{J}}}_\star$ and $\bds{\hat{h}}$  were adopted, while in the upper right panel they were increased. In both upper panels, a small offset of $5^\circ$ between the initial orientations of $\bds{\hat{S}}$ and ${\bds{\hat{J}}}_\star$ was introduced. In the lower left panel, an initial spin-orbit misalignment as little as $\eta_\star=5^\circ$ was assumed, while ${\bds{\hat{S}}}_0$ was allowed to differ sensibly with respect to ${\bds{\hat{J}}}_\star$. In the lower right panel, $\eta_\star=50^\circ$ and huge differences in the initial orientations between ${\bds{\hat{S}}}_0$ and ${\bds{\hat{J}}}_\star$ were assumed.}\label{figura1}
\end{figure*}

\begin{figure*}[ht]
\centering
\begin{tabular}{cc}
\includegraphics[width=8cm]{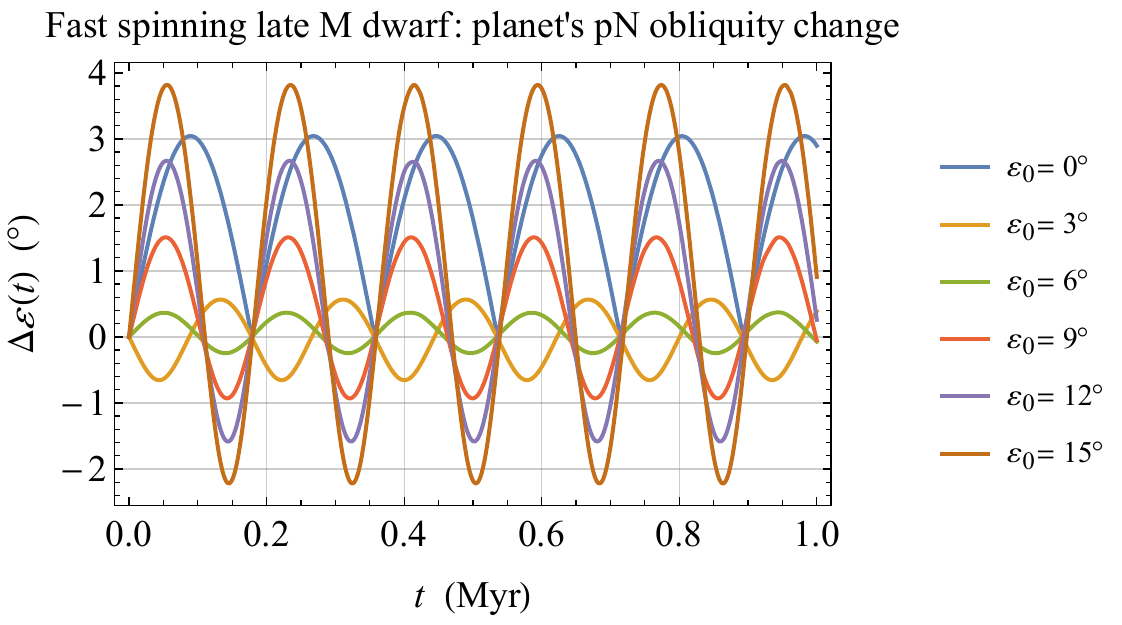} & \includegraphics[width=8cm]{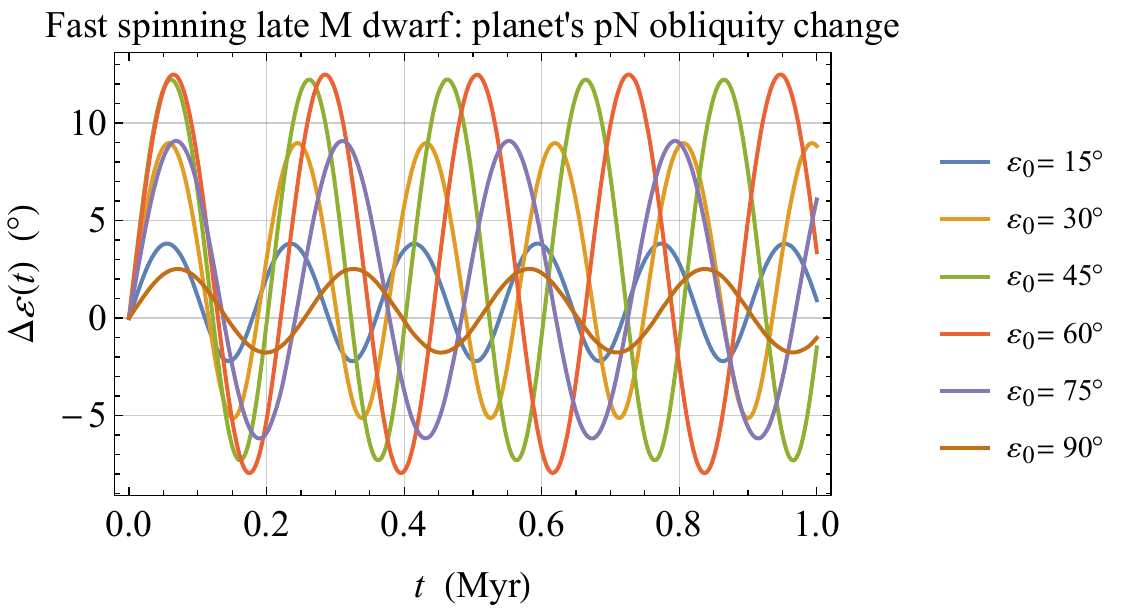}\\
\includegraphics[width=8cm]{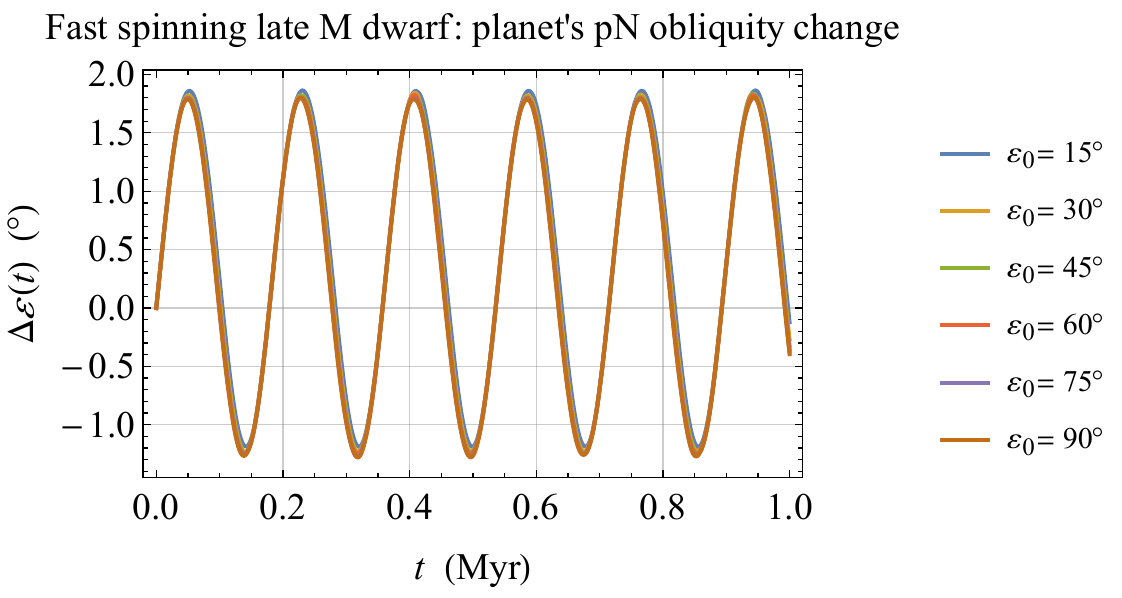} & \includegraphics[width=8cm]{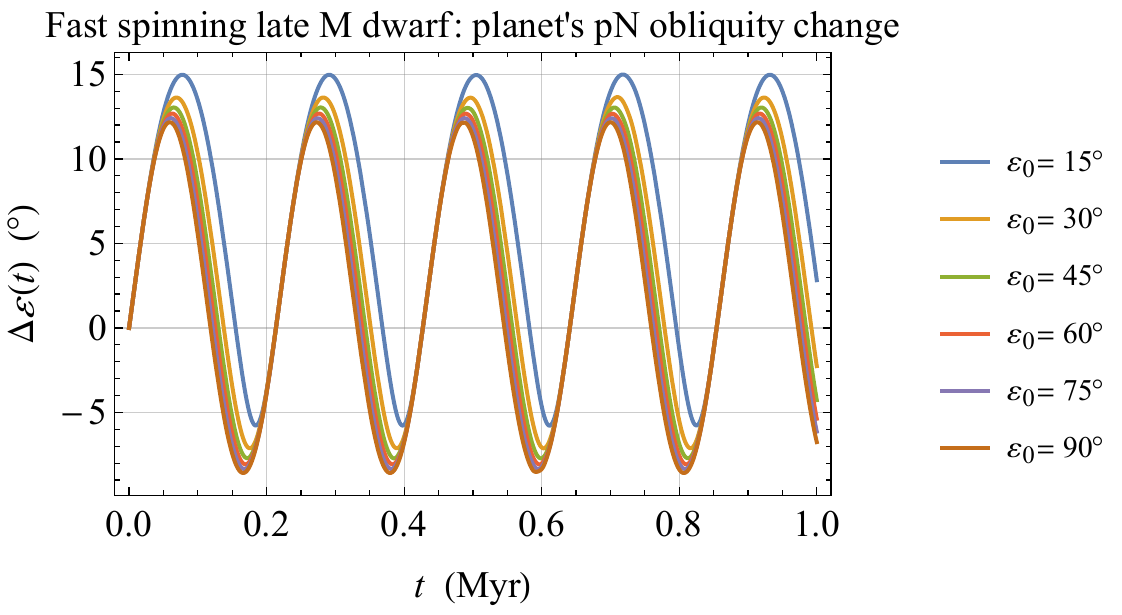}\\
\end{tabular}
\caption{
Numerically integrated time series of the \textcolor{black}{$J_2^\star$-driven pK} variation $\Delta\varepsilon(t)$ of the time-dependent obliquity $\varepsilon(t)$ of the spin axis $\bds{\hat{S}}$ of a fictitious potentially habitable planet orbiting a fast spinning late-type M dwarf characterized by $M_\star=0.089\,M_\odot,\,R_\star=0.107\,R_\odot,\,k_2^\star=0.16,\,P_\star = 0.5\,\mathrm{d}$ and different initial orientations of the planetary and stellar spin angular momenta $\bds{\hat{S}},\,\bds{\hat{J}}_\star$, all referred to the ecliptic plane at the initial epoch as reference $\grf{x,\,y}$ plane. \Rfr{oblateness} was used after integrating \textcolor{black}{\rfr{dLdtJ2}}. The initial values adopted for the relevant spin and orbital parameters, common to all the integrations, are $e=0.0,\,a=0.0252\,\mathrm{au},\,I_0=\textcolor{black}{0.0^\circ},\,\Omega_0=50^\circ,\,\alpha_0=50^\circ,\,\varphi_\star=150^\circ$.  In the upper left panel, moderate spin-orbit misalignments between the initial configurations of ${\bds{\hat{J}}}_\star$ and $\bds{\hat{h}}$  were adopted, while in the upper right panel they were increased. In both upper panels, a small offset of $5^\circ$ between the initial orientations of $\bds{\hat{S}}$ and ${\bds{\hat{J}}}_\star$ was introduced. In the lower left panel, an initial spin-orbit misalignment as little as $\eta_\star=5^\circ$ was assumed, while ${\bds{\hat{S}}}_0$ was allowed to differ sensibly with respect to ${\bds{\hat{J}}}_\star$. In the lower right panel, $\eta_\star=50^\circ$ and huge differences in the initial orientations between ${\bds{\hat{S}}}_0$ and ${\bds{\hat{J}}}_\star$ were assumed.}\label{figura2}
\end{figure*}

\begin{figure*}[ht]
\centering
\begin{tabular}{cc}
\includegraphics[width=8cm]{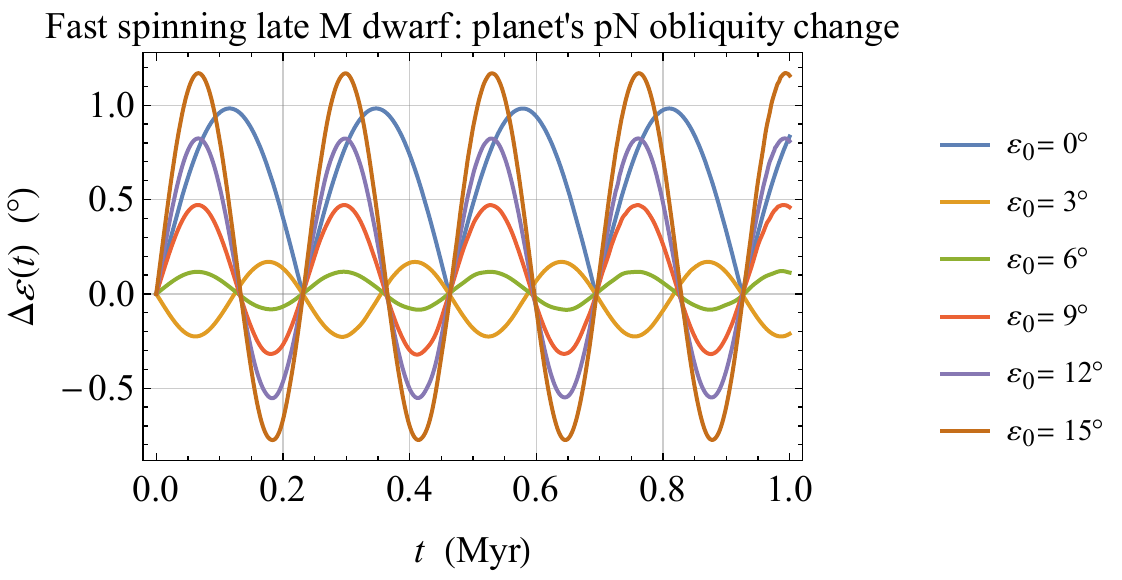} & \includegraphics[width=8cm]{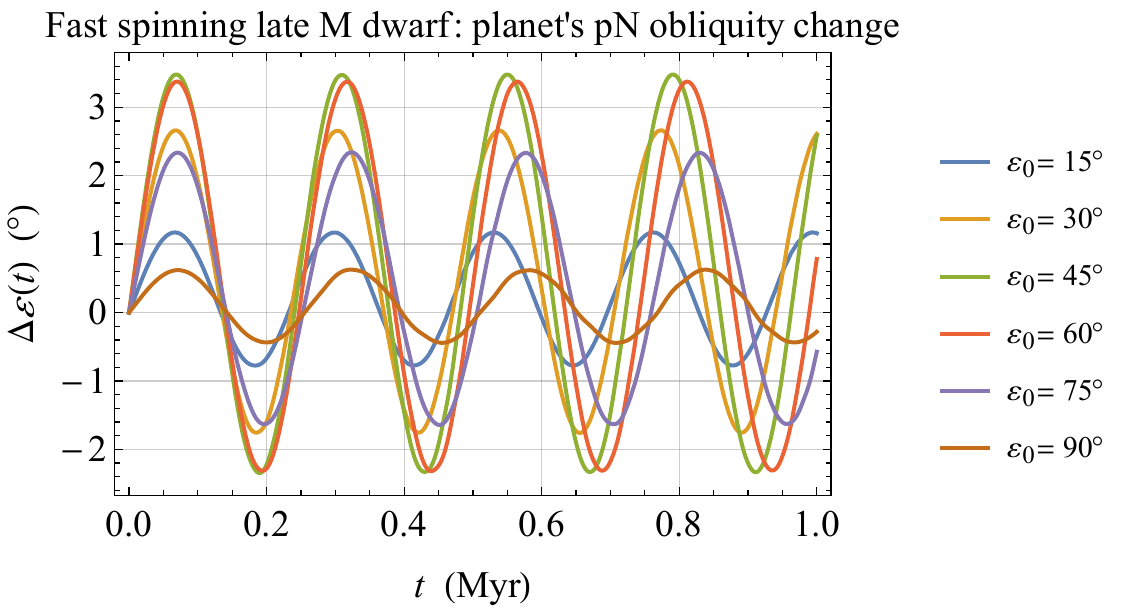}\\
\includegraphics[width=8cm]{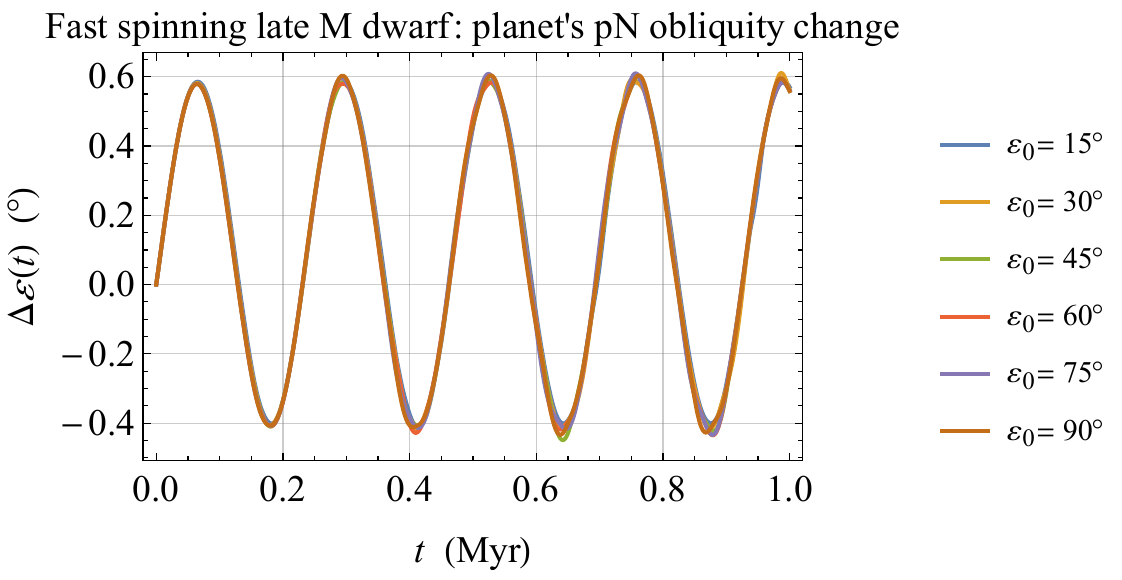} & \includegraphics[width=8cm]{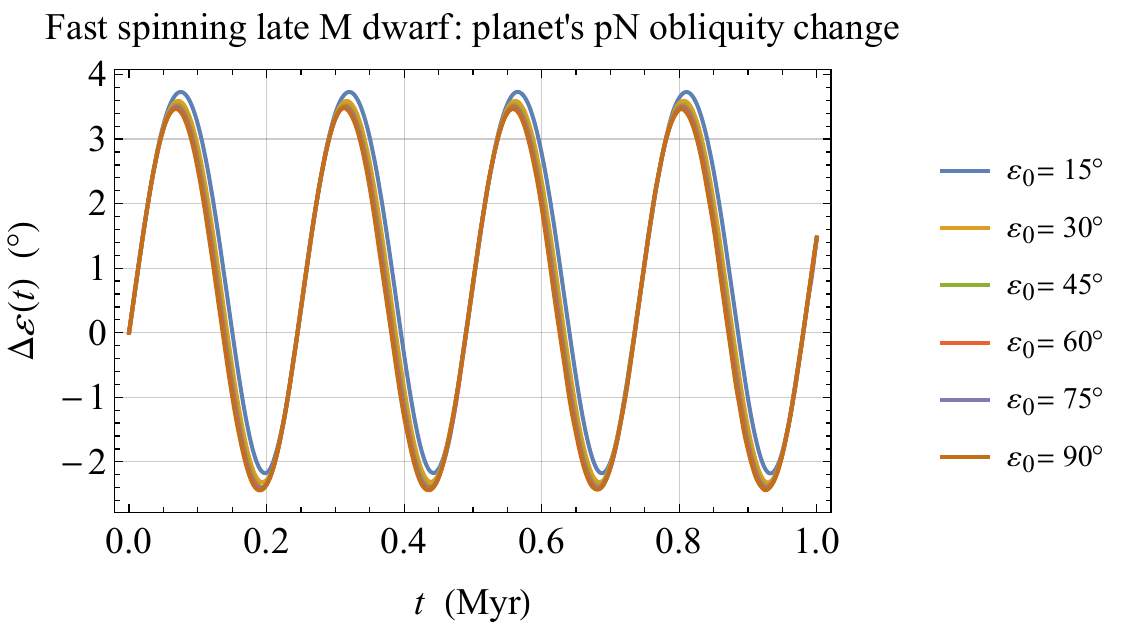}\\
\end{tabular}
\caption{
Numerically integrated time series of the \textcolor{black}{$J_2^\star$-driven pK} variation $\Delta\varepsilon(t)$ of the time-dependent obliquity $\varepsilon(t)$ of the spin axis $\bds{\hat{S}}$ of a fictitious potentially habitable planet orbiting a fast spinning late-type M dwarf characterized by $M_\star=0.089\,M_\odot,\,R_\star=0.107\,R_\odot,\,k_2^\star=0.16,\,P_\star = 1\,\mathrm{d}$ and different initial orientations of the planetary and stellar spin angular momenta $\bds{\hat{S}},\,\bds{\hat{J}}_\star$, all referred to the ecliptic plane at the initial epoch as reference $\grf{x,\,y}$ plane. \Rfr{oblateness} was used after integrating \textcolor{black}{\rfr{dLdtJ2}}. The initial values adopted for the relevant spin and orbital parameters, common to all the integrations, are $e=0.0,\,a=0.0252\,\mathrm{au},\,I_0=\textcolor{black}{0.0^\circ},\,\Omega_0=50^\circ,\,\alpha_0=50^\circ,\,\varphi_\star=150^\circ$.  In the upper left panel, moderate spin-orbit misalignments between the initial configurations of ${\bds{\hat{J}}}_\star$ and $\bds{\hat{h}}$  were adopted, while in the upper right panel they were increased. In both upper panels, a small offset of $5^\circ$ between the initial orientations of $\bds{\hat{S}}$ and ${\bds{\hat{J}}}_\star$ was introduced. In the lower left panel, an initial spin-orbit misalignment as little as $\eta_\star=5^\circ$ was assumed, while ${\bds{\hat{S}}}_0$ was allowed to differ sensibly with respect to ${\bds{\hat{J}}}_\star$. In the lower right panel, $\eta_\star=50^\circ$ and huge differences in the initial orientations between ${\bds{\hat{S}}}_0$ and ${\bds{\hat{J}}}_\star$ were assumed.}\label{figura3}
\end{figure*}

\lb{sec.4}
In \figs{figura1}{figura3}, I display the numerically produced time series of the \textcolor{black}{oblateness-driven classical pK} variation $\Delta\varepsilon(t)$ of the time-dependent obliquity $\varepsilon(t)$ of the spin axis $\bds{\hat{S}}$ of a planet orbiting at $r=0.0252\,\mathrm{au}$ about a fast spinning late-type M dwarf with $M_\star=0.089\,M_\odot,\,R_\star=0.107\,R_\odot,\,k_2^\star=0.16,\,\varphi_\star = 150^\circ$ for various star's spinning periods: $P_\star=0.1\,\mathrm{d}$ (\fig\ref{figura1}), $P_\star=0.5\,\mathrm{d}$ (\fig\ref{figura2}), and $P_\star=1\,\mathrm{d}$ (\fig\ref{figura3}). In each figure, I obtained them by means of \rfr{oblateness} after having numerically integrated\textcolor{black}{\footnote{\textcolor{black}{For the sake of completeness, I included also \rfr{dJdt}  in the integration by assuming $M=1.05\,M_\oplus$, but it did not have noticeable effects on the resulting patterns of \figs{figura1}{figura3}.}}} \textcolor{black}{\rfr{dLdtJ2} and \rfr{J2Pstar}  for $J_2^\star$}  over a time span $\Delta t = 0.1-1\,\mathrm{Myr}$.  I assumed the ecliptic plane at the initial epoch as reference $\grf{x,\,y}$ plane\footnote{It turned out that $\Omega_0$ does not impact the patterns of $\Delta\varepsilon\ton{t}$; thus, I assumed $\Omega_0=50^\circ$. } ($I_0\textcolor{black}{=} 0^\circ,\,\theta_0\textcolor{black}{=}\varepsilon_0$), and I varied  the initial values of the tilt of $\bds{\hat{S}}$ and ${\bds{\hat{J}}}_\star$ in each run by keeping, say, $\alpha_0=50^\circ$.

In the upper rows, I introduced a small offset $\delta = -5^\circ$ between $\varepsilon_0$ and $\eta_\star=\varepsilon_0+\delta$, while I removed such a limitation in the lower rows. In the left panels, a moderate spin-orbit misalignment  between ${\bds{\hat{J}}}_\star$ and $\bds{\hat{h}}$ was assumed, while it was increased in the right panels.

From \fig\ref{figura1}, it can be noted that, for $P_\star = 0.1\,\mathrm{d}$, the obliquity $\varepsilon$ may experience significative variations, as large as tens of degrees, with characteristic time scales as short as $\simeq 20\,\mathrm{kyr}$. They are more evident for large values of the spin-orbit misalignment between ${\bds{\hat{J}}}_\star$ and $\bds{\hat{h}}$, as shown by the upper and lower right panels, where the peak-to-peak obliquity variations can reach $\lesssim 100-120^\circ$. They are of the order of $\lesssim 10-20^\circ$ if the star's spin axis is almost perpendicular to the ecliptic plane (upper and lower left panels).

\fig\ref{figura2} shows that, for $P_\star = 0.5\,\mathrm{d}$, the magnitude of the obliquity variations reduces down to $\lesssim 3-6^\circ$ for ${\bds{\hat{J}}}_\star$ and $\bds{\hat{h}}$ almost aligned (upper and lower left panels), while it can still be as large as $\lesssim 20-25^\circ$ for a tilted stellar spin axis (upper and lower right panels). The characteristic timescale is $\simeq 0.2\,\mathrm{Myr}$.

In \fig\ref{figura3}, the case for $P_\star = 1\,\mathrm{d}$ is depicted. From the upper and lower left panels, it turns out that the magnitude of the obliquity variations drops to $\lesssim 1-2^\circ$ if the stellar spin axis is almost perpendicular to the ecliptic plane. Instead, as shown by the upper and lower right panels, the peak-to-peak amplitudes of $\Delta\varepsilon$ are $\lesssim 6-7^\circ$ if ${\bds{\hat{J}}}_\star$ and $\bds{\hat{h}}$ are misaligned. The characteristic time scale is somewhat longer than $0.2\,\mathrm{Myr}$.

I displayed a necessarily limited part of the parameter space; it can be shown that the same features hold also by varying $\alpha_0$ and $\varphi_\star$. Just as an example, by varying $\alpha_0$ from $30^\circ$ to $340^\circ$ with $\varepsilon_0=23\grd 44,\,\eta_\star = \varepsilon_0 + 5^\circ,\,\varphi_\star=150^\circ$, I get $\Delta\varepsilon\lesssim 70^\circ,\,\ton{P_\star = 0.1\,\mathrm{d}},\,\Delta\varepsilon\lesssim 30^\circ,\,\ton{P_\star = 0.5\,\mathrm{d}},\,\Delta\varepsilon\lesssim 10^\circ,\,\ton{P_\star = 1\,\mathrm{d}}$. Essentially the same occurs by varying $\varphi_\star$ from $60^\circ$ to $360^\circ$ with $\varepsilon_0=23\grd 44,\,\alpha_0 = 50^\circ,\,\eta_\star = \varepsilon_0 + 5^\circ$.

\begin{figure*}[ht]
\centering
\begin{tabular}{cc}
\includegraphics[width=8cm]{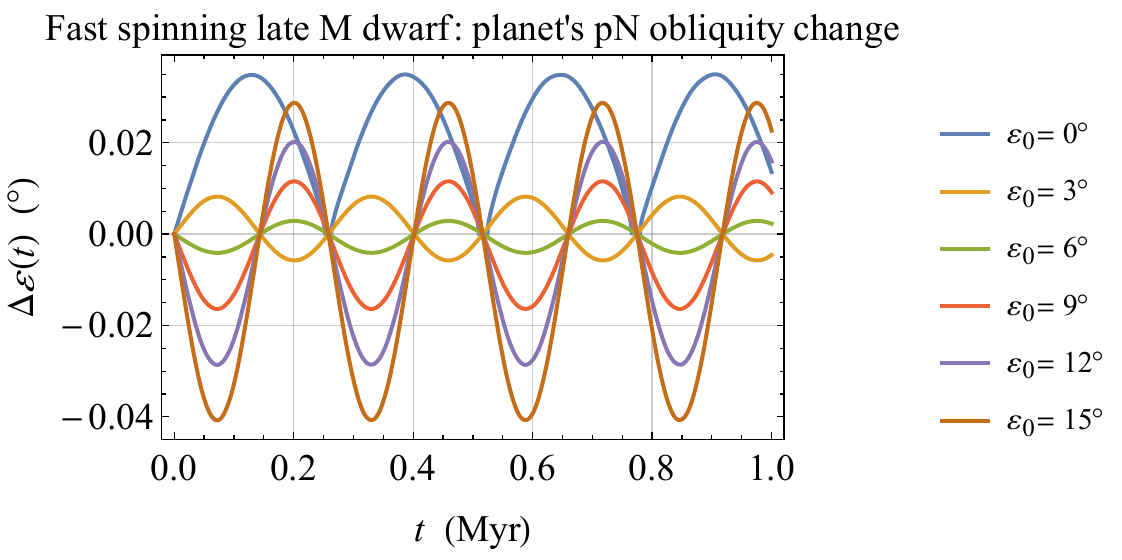} & \includegraphics[width=8cm]{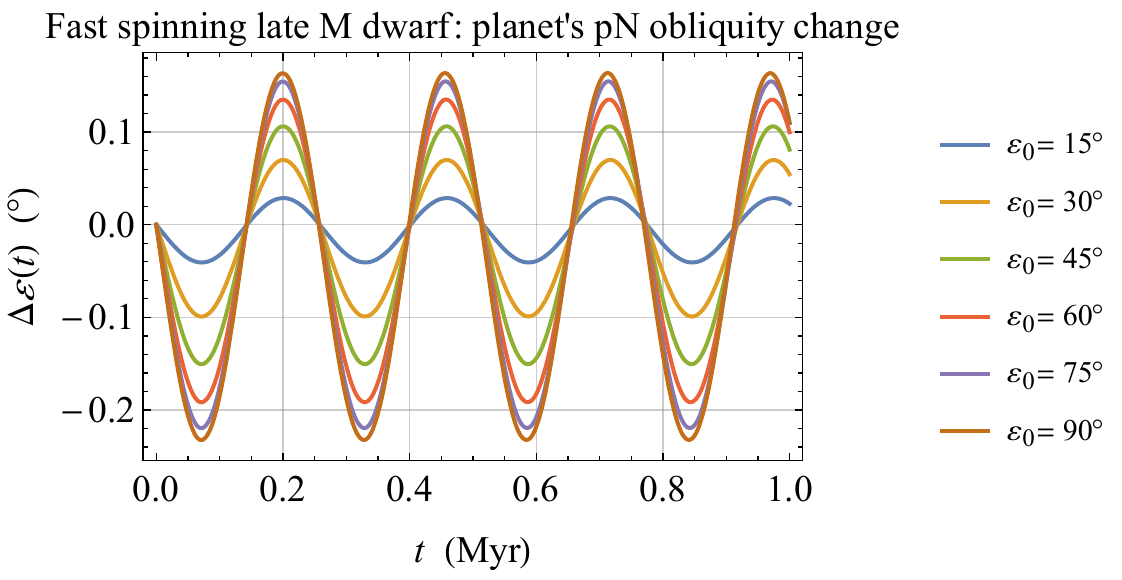}\\
\includegraphics[width=8cm]{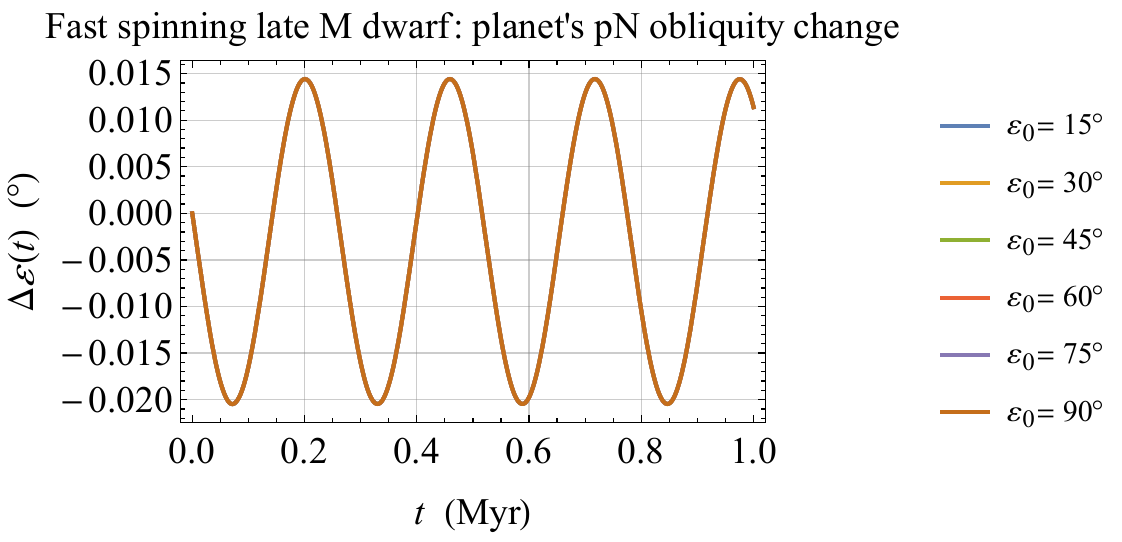} & \includegraphics[width=8cm]{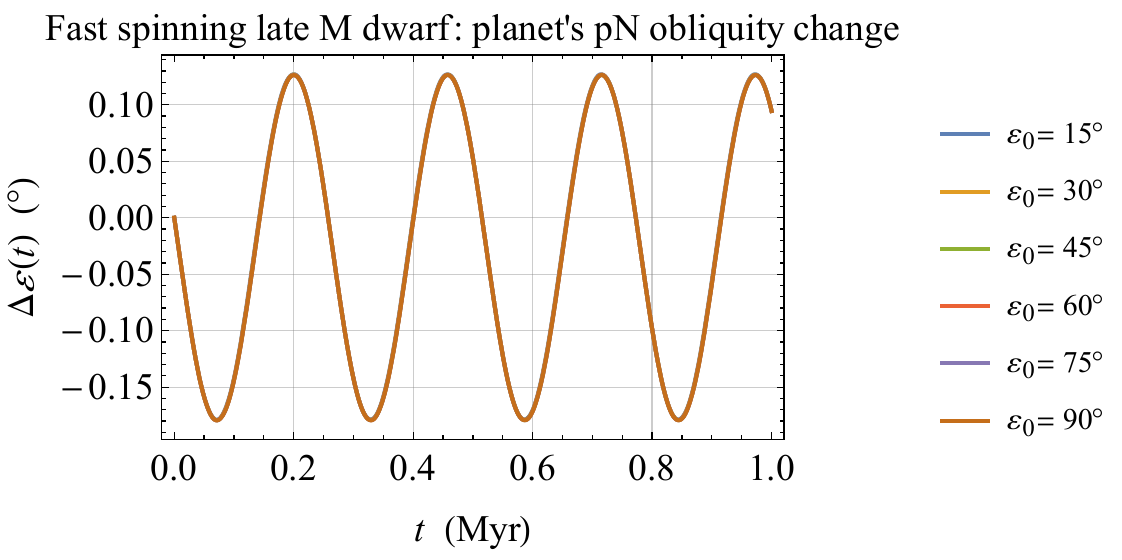}\\
\end{tabular}
\caption{
\textcolor{black}{Numerically integrated time series of the purely pN variation $\Delta\varepsilon(t)$ of the time-dependent obliquity $\varepsilon(t)$ of the spin axis $\bds{\hat{S}}$ of a fictitious potentially habitable planet orbiting a fast spinning late-type M dwarf characterized by $M_\star=0.089\,M_\odot,\,R_\star=0.107\,R_\odot,\,k_2^\star=0.16,\,P_\star = 0.1\,\mathrm{d}$ and different initial orientations of the planetary and stellar spin angular momenta $\bds{\hat{S}},\,\bds{\hat{J}}_\star$, all referred to the ecliptic plane at the initial epoch as reference $\grf{x,\,y}$ plane. \Rfr{oblateness} was used after simultaneously integrating \rfr{dSdtpN} and \rfr{dLdtpN}. The initial values adopted for the relevant spin and orbital parameters, common to all the integrations, are $e=0.0,\,a=0.0252\,\mathrm{au},\,I_0=0.0^\circ,\,\Omega_0=50^\circ,\,\alpha_0=50^\circ,\,\varphi_\star=150^\circ$.  In the upper left panel, moderate spin-orbit misalignments between the initial configurations of ${\bds{\hat{J}}}_\star$ and $\bds{\hat{h}}$  were adopted, while in the upper right panel they were increased. In both upper panels, a small offset of $5^\circ$ between the initial orientations of $\bds{\hat{S}}$ and ${\bds{\hat{J}}}_\star$ was introduced. In the lower left panel, an initial spin-orbit misalignment as little as $\eta_\star=5^\circ$ was assumed, while ${\bds{\hat{S}}}_0$ was allowed to differ sensibly with respect to ${\bds{\hat{J}}}_\star$. In the lower right panel, $\eta_\star=50^\circ$ and huge differences in the initial orientations between ${\bds{\hat{S}}}_0$ and ${\bds{\hat{J}}}_\star$ were assumed.}}\label{figura1bis}
\end{figure*}

\textcolor{black}{Figure\,\ref{figura1bis} was obtained by simultaneously integrating the pN \rfr{dSdtpN} and \rfr{dLdtpN}, being all the orbital, spin and physical parameters of the star-planet binary identical to those of Figure\,\ref{figura1}; the stellar oblateness-driven classical precession of the orbital plane due to \rfr{dLdtJ2} was not included.
}
\textcolor{black}{The resulting purely pN signatures are negligible, amounting to less that $\Delta\varepsilon\ll 1^\circ$.}

\begin{figure*}[ht]
\centering
\begin{tabular}{cc}
\includegraphics[width=8cm]{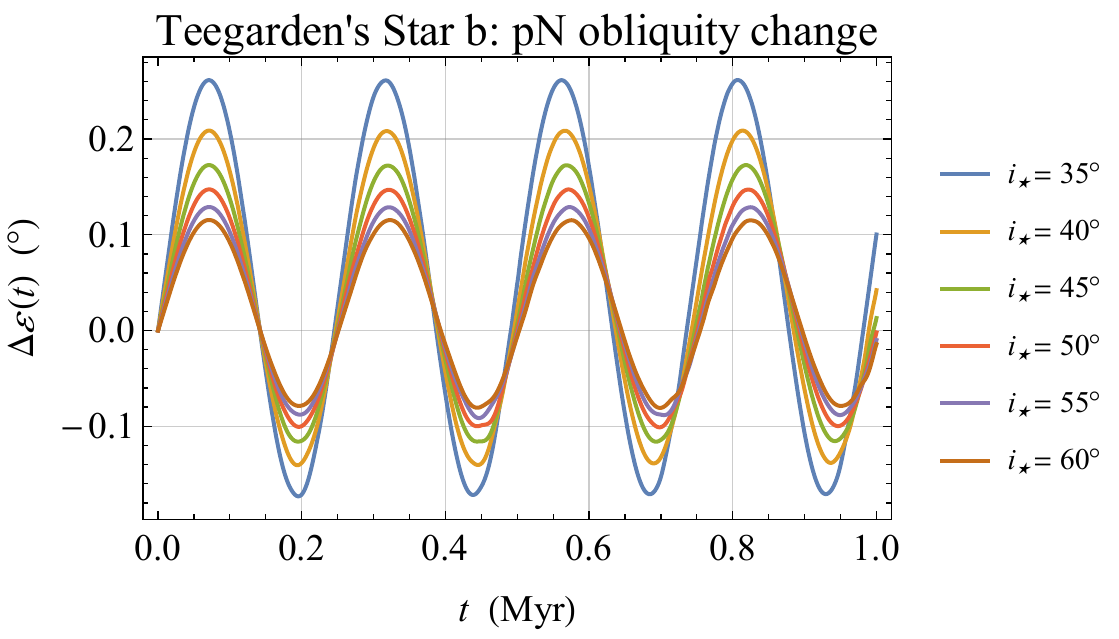} & \includegraphics[width=8cm]{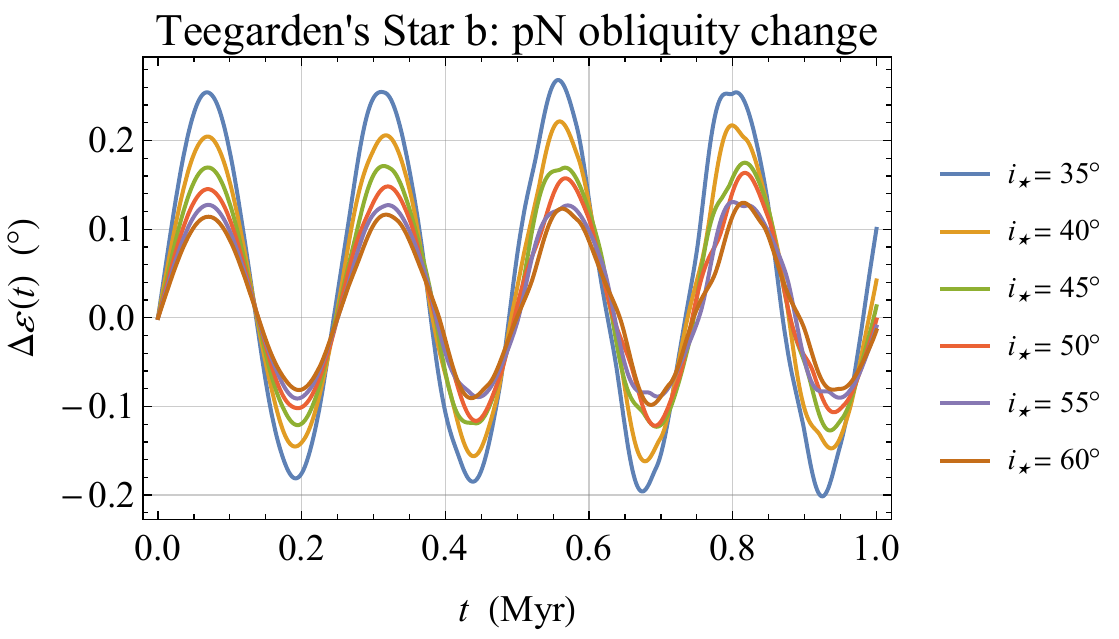}\\
\includegraphics[width=8cm]{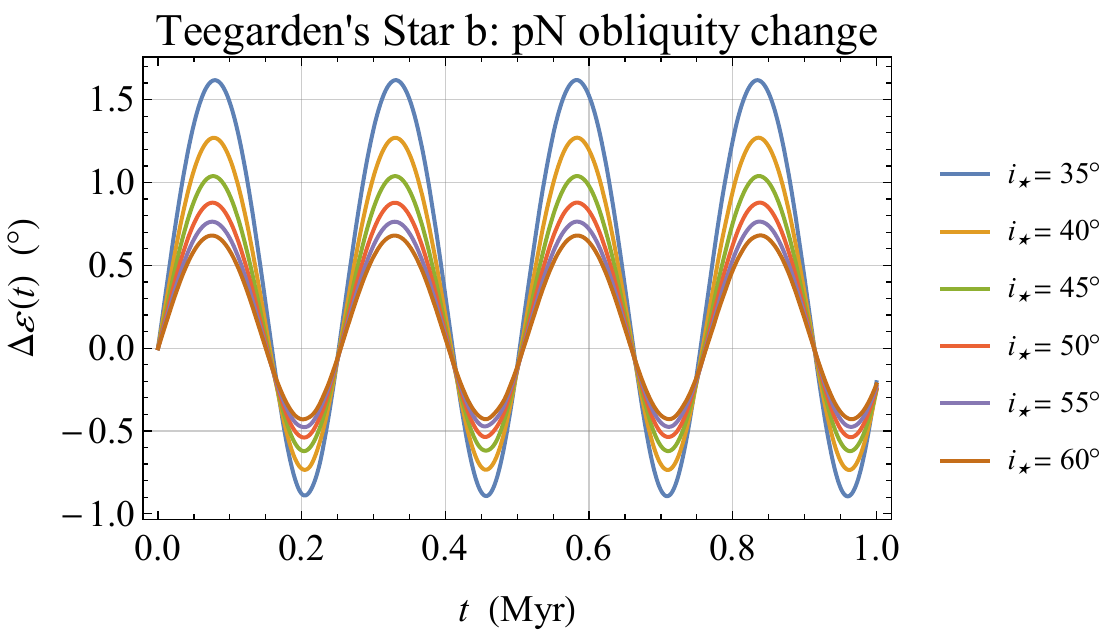} & \includegraphics[width=8cm]{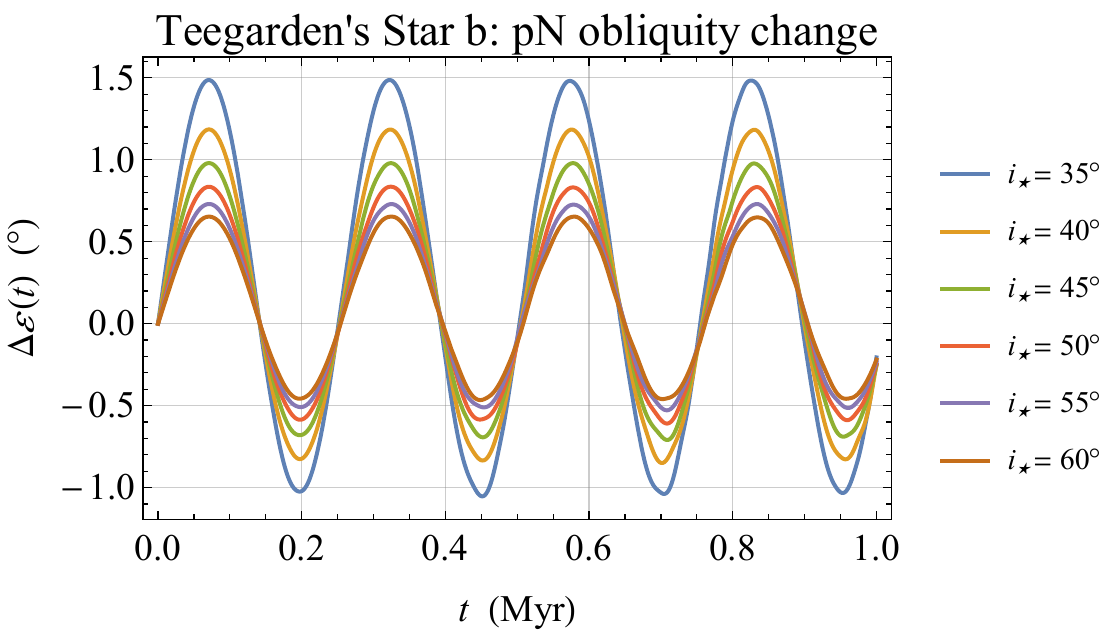}\\
\end{tabular}
\caption{
Numerically integrated time series of the \textcolor{black}{$J_2^\star$-driven pK} variation $\Delta\varepsilon(t)$ of the time-dependent obliquity $\varepsilon(t)$ of the spin axis $\bds{\hat{S}}$ of the planet b \citep{2019A&A...627A..49Z} of the Teegarden's Star \citep{2003ApJ...589L..51T} with $M_\star=0.089\,M_\odot,\,R_\star=0.107\,R_\odot,\,k_2^\star=0.16,\,u_\star = 2\,\mathrm{km\,s}^{-1}$ and different initial orientations of the planetary and stellar spin angular momenta $\bds{\hat{S}},\,\bds{\hat{J}}_\star$, all referred to the ecliptic plane at the initial epoch as reference $\grf{x,\,y}$ plane. \Rfr{oblateness} was used after integrating \textcolor{black}{\rfr{dLdtJ2}}.  The initial values adopted for the relevant spin and orbital parameters, common to all the integrations, are $e=0.0,\,a=0.0252\,\mathrm{au},\,I_0=\textcolor{black}{0.0^\circ},\,\Omega_0=50^\circ,\,\alpha_0=50^\circ,\,\varphi_\star=150^\circ$.  In each panel, $i_\star$ is varied in the range $35^\circ\leq i_\star\leq 60^\circ$. In the upper left panel, $\bds{\hat{S}},\,\bds{\hat{J}}_\star,\,\bds{\hat{h}}$ are initially almost aligned ($\varepsilon_0=3^\circ,\,\eta_\star=5^\circ$). In the upper right panel, the initial spin-orbit misalignment between $\bds{\hat{J}}_\star$ and $\bds{\hat{h}}$ is modest ($\eta_\star=5^\circ$), while $\bds{\hat{S}}_0$ is notably tilted ($\varepsilon_0 = 50^\circ$). In the lower left panel, $\bds{\hat{S}}$ and $\bds{\hat{h}}$ are initially aligned ($\varepsilon_0=5^\circ)$, while $\bds{\hat{J}}_\star$ is strongly tilted ($\eta_\star=50^\circ$). In the lower right panel, $\bds{\hat{S}}$ and $\bds{\hat{J}}_\star$ are initially almost aligned and remarkably tilted ($\varepsilon_0 = 45^\circ,\,\eta_\star = 50^\circ$).}\label{figura4}
\end{figure*}

In the case of the planet b \citep{2019A&A...627A..49Z} of the Teegarden's Star \citep{2003ApJ...589L..51T}, no photometric rotation period is available for the M dwarf, whose projected rotation velocity $u_\star$ seems to be no larger than $u_\star\lesssim 2\,\mathrm{km\,s}^{-1}$ \citep{2018A&A...612A..49R} because of the absence of any significant rotational broadening in its spectroscopic measurements; $i_\star$ is unknown. Thus, I will use \rfr{J2q} and \rfr{Jstar} for $J_2^\star$ and $J_\star$, respectively. In \fig\ref{figura4}, the result of my numerical integrations of \textcolor{black}{\rfr{dSdtpN} and \rfr{dLdt}}, performed by assuming $u_\star = 2\,\mathrm{km\,s}^{-1}$, and of the resulting calculation of the time series for \rfr{oblateness} is displayed.
I varied $i_\star$ from a run to another, and all the runs in each panel are characterized by a different initial orientation of $\bds{\hat{S}}$ and $\bds{\hat{J}}_\star$. It turns out that, in all the cases examined, the amplitude of the pN shift $\Delta\varepsilon(t)$ is essentially negligible, amounting to about less than a degree or so with a characteristic time scale of $0.2\,\mathrm{Myr}$. Thus, it can be concluded that it is unlikely that the pN de Sitter and Pugh-Schiff planetary spin precessions may play a significant role in the habitability of the planet b of the Teegarden's Star.
%
%
\section{Summary and conclusions}\lb{sec.5}
\textcolor{black}{The oblateness of low mass, fast spinning late M dwarfs of spectral class M9V} can play a relevant role\textcolor{black}{, to the Newtonian level,} in the overall assessment of the habitability of potentially habitable telluric planets \textcolor{black}{orbiting them}. \textcolor{black}{Instead, the effects of GTR are negligible.}

Indeed, the resulting long-term \textcolor{black}{pK} variations of the obliquity $\varepsilon$ of the spin $\bds S$ of a fictitious Earth-like planet orbiting a $M=0.08\,M_\odot$ late-type M dwarf in $\Pb\simeq 4\,\mathrm{d}$ have peak-to-peak amplitudes $\Delta\varepsilon$ which can be as large as tens of degrees for $P_\star \simeq 0.1-0.5\,\mathrm{d}$, with characteristic timescales of the order of $\simeq 20-200\,\mathrm{kyr}$. Large values of the spin-orbit misalignment between the star's spin ${\bds J}_\star$ and the orbital angular momentum $\bds L$ tend to strengthen the size of the \textcolor{black}{pK} obliquity variations. For $P_\star=1\,\mathrm{d}$, if the star's spin is tilted to the ecliptic plane, the magnitude of the \textcolor{black}{pK} effect falls down to $\Delta\varepsilon\lesssim 6-7^\circ$ with a periodicity of $200\,\mathrm{kyr}$.

The spin of the existing planet b of the slowly rotating Teergarden's Star is displaced by just $\Delta\varepsilon\lesssim 1-1.5^\circ$, with a periodicity as long as $200\,\mathrm{kyr}$.

\textcolor{black}{As directions for future work, it is worth mentioning the investigation of the impact on $\varepsilon$ of the oblateness $J_2$ of the planet itself and of its tidal distortion.
}

The approach presented here is not limited just to extrasolar planets, being suitable to be extended also to other astronomical and astrophysical scenarios whose primary interest is not the habitability of the target bodies.
\section*{\textcolor{black}{Acknowledgements}}
\textcolor{black}{I am grateful to an anonymous referee for her/his critical remarks which improved the manuscript.}
\bibliographystyle{aa} 
\bibliography{exopbib,PXbib}{}

\begin{thebibliography}{113}
\expandafter\ifx\csname natexlab\endcsname\relax\def\natexlab#1{#1}\fi

\bibitem[{{Adams}, {Bodenheimer} \& {Laughlin}(2005){Adams}, {Bodenheimer}, \&
  {Laughlin}}]{2005AN....326..913A}
{Adams} F.~C., {Bodenheimer} P., {Laughlin} G., 2005, AN, 326, 913

\bibitem[{{Armstrong} {et~al}\mbox{.}(2014){Armstrong}, {Barnes},
  {Domagal-Goldman}, {Breiner}, {Quinn}, \& {Meadows}}]{2014AsBio..14..277A}
{Armstrong} J.~C., {Barnes} R., {Domagal-Goldman} S., {Breiner} J., {Quinn}
  T.~R., {Meadows} V.~S., 2014, AsBio, 14, 277

\bibitem[{{Armstrong}, {Leovy} \& {Quinn}(2004){Armstrong}, {Leovy}, \&
  {Quinn}}]{2004Icar..171..255A}
{Armstrong} J.~C., {Leovy} C.~B., {Quinn} T., 2004, Icar, 171, 255

\bibitem[{{Asada} \& {Futamase}(1997)}]{1997PThPS.128..123A}
{Asada} H., {Futamase} T., 1997, PThPS, 128, 123

\bibitem[{Awramik(1992)}]{Aw92}
Awramik S., 1992, Photosynth. Res., 33, 75

\bibitem[{{Barker} \& {O'Connell}(1975)}]{1975PhRvD..12..329B}
{Barker} B.~M., {O'Connell} R.~F., 1975, PhRvD, 12, 329

\bibitem[{{Barnes} {et~al}\mbox{.}(2015){Barnes}, {Jeffers}, {Jones},
  {Pavlenko}, {Jenkins}, {Haswell}, \& {Lohr}}]{2015ApJ...812...42B}
{Barnes} J.~R., {Jeffers} S.~V., {Jones} H.~R.~A., {Pavlenko} Y.~V., {Jenkins}
  J.~S., {Haswell} C.~A., {Lohr} M.~E., 2015, \apj, 812, 42

\bibitem[{{Barnes}(2009)}]{2009ApJ...705..683B}
{Barnes} J.~W., 2009, \apj, 705, 683

\bibitem[{{Becker} {et~al}\mbox{.}(2018){Becker}, {Bethkenhagen}, {Kellermann},
  {Wicht}, \& {Redmer}}]{2018AJ....156..149B}
{Becker} A., {Bethkenhagen} M., {Kellermann} C., {Wicht} J., {Redmer} R., 2018,
  \aj, 156, 149

\bibitem[{Bell {et~al}\mbox{.}(2015)Bell, Boehnke, Harrison, \&
  Mao}]{Bell14518}
Bell E.~A., Boehnke P., Harrison T.~M., Mao W.~L., 2015, PNAS, 112, 14518

\bibitem[{{Bessell}(1991)}]{1991AJ....101..662B}
{Bessell} M.~S., 1991, \aj, 101, 662

\bibitem[{{Birkby}(2018)}]{2018haex.bookE..16B}
{Birkby} J.~L., 2018, in Handbook of Exoplanets, {Deeg} H.~J., {Belmonte}
  J.~A., eds., p.~16

\bibitem[{{Blanchet}(2003)}]{2003Blanchet}
{Blanchet} L., 2003, in Proceedings of the Twelfth Workshop on General
  relativity and Gravitation in Japan, {Shibata} M., {Eriguchi} Y., {Taniguchi}
  K., {Nakamura} T., {Tomita} K., eds., The University of Tokyo, Komaba, Tokyo,
  pp. 8--23

\bibitem[{{Bourda} \& {Capitaine}(2004)}]{2004A&A...428..691B}
{Bourda} G., {Capitaine} N., 2004, \aap, 428, 691

\bibitem[{{Brandt} {et~al}\mbox{.}(2020){Brandt}, {Dupuy}, {Bowler}, {Bardalez
  Gagliuffi}, {Faherty}, {Brandt}, \& {Michalik}}]{2020AJ....160..196B}
{Brandt} T.~D., {Dupuy} T.~J., {Bowler} B.~P., {Bardalez Gagliuffi} D.~C.,
  {Faherty} J., {Brandt} G.~M., {Michalik} D., 2020, \aj, 160, 196

\bibitem[{{Breton} {et~al}\mbox{.}(2008){Breton}, {Kaspi}, {Kramer},
  {McLaughlin}, {Lyutikov}, {Ransom}, {Stairs}, {Ferdman}, {Camilo}, \&
  {Possenti}}]{2008Sci...321..104B}
{Breton} R.~P. {et~al.}, 2008, Sci, 321, 104

\bibitem[{{Chabrier} {et~al}\mbox{.}(2000){Chabrier}, {Baraffe}, {Allard}, \&
  {Hauschildt}}]{2000ApJ...542..464C}
{Chabrier} G., {Baraffe} I., {Allard} F., {Hauschildt} P., 2000, \apj, 542, 464

\bibitem[{{Claret}(2004)}]{2004A&A...424..919C}
{Claret} A., 2004, \aap, 424, 919

\bibitem[{{Correia} {et~al}\mbox{.}(2011){Correia}, {Laskar}, {Farago}, \&
  {Bou{\'e}}}]{2011CeMDA.111..105C}
{Correia} A. C.~M., {Laskar} J., {Farago} F., {Bou{\'e}} G., 2011, CeMDA, 111,
  105

\bibitem[{{de Sitter}(1916)}]{1916MNRAS..77..155D}
{de Sitter} W., 1916, MNRAS, 77, 155

\bibitem[{{Debono} \& {Smoot}(2016)}]{2016Univ....2...23D}
{Debono} I., {Smoot} G.~F., 2016, Univ, 2, 23

\bibitem[{{Deeg} \& {Alonso}(2018)}]{2018haex.bookE.117D}
{Deeg} H.~J., {Alonso} R., 2018, in Handbook of Exoplanets, {Deeg} H.~J.,
  {Belmonte} J.~A., eds., p. 117

\bibitem[{{Deeg} \& {Belmonte}(2018)}]{2018haex.bookE....D}
{Deeg} H.~J., {Belmonte} J.~A., 2018, {Handbook of Exoplanets}. Springer, Cham

\bibitem[{{Deleuil} {et~al}\mbox{.}(2008){Deleuil}, {Deeg}, {Alonso}, {Bouchy},
  {Rouan}, {Auvergne}, {Baglin}, {Aigrain}, {Almenara}, {Barbieri}, {Barge},
  {Bruntt}, {Bord{\'e}}, {Collier Cameron}, {Csizmadia}, {de La Reza},
  {Dvorak}, {Erikson}, {Fridlund}, {Gandolfi}, {Gillon}, {Guenther}, {Guillot},
  {Hatzes}, {H{\'e}brard}, {Jorda}, {Lammer}, {L{\'e}ger}, {Llebaria},
  {Loeillet}, {Mayor}, {Mazeh}, {Moutou}, {Ollivier}, {P{\"a}tzold}, {Pont},
  {Queloz}, {Rauer}, {Schneider}, {Shporer}, {Wuchterl}, \&
  {Zucker}}]{2008A&A...491..889D}
{Deleuil} M. {et~al.}, 2008, \aap, 491, 889

\bibitem[{{Dickey} {et~al}\mbox{.}(1994){Dickey}, {Bender}, {Faller},
  {Newhall}, {Ricklefs}, {Ries}, {Shelus}, {Veillet}, {Whipple}, {Wiant},
  {Williams}, \& {Yoder}}]{1994Sci...265..482D}
{Dickey} J.~O. {et~al.}, 1994, Sci, 265, 482

\bibitem[{{Dobrovolskis}(2013)}]{2013Icar..226..760D}
{Dobrovolskis} A.~R., 2013, Icar, 226, 760

\bibitem[{{Dole}(1964)}]{1964hpfm.book.....D}
{Dole} S.~H., 1964, {Habitable planets for man}. Blaisdell Publishing, New York

\bibitem[{{Dong}, {Huang} \& {Lingam}(2019){Dong}, {Huang}, \&
  {Lingam}}]{2019ApJ...882L..16D}
{Dong} C., {Huang} Z., {Lingam} M., 2019, \apjl, 882, L16

\bibitem[{{Engle} \& {Guinan}(2018)}]{2018RNAAS...2...34E}
{Engle} S.~G., {Guinan} E.~F., 2018, RNAAS, 2, 34

\bibitem[{{Everitt} {et~al}\mbox{.}(2011){Everitt}, {Debra}, {Parkinson},
  {Turneaure}, {Conklin}, {Heifetz}, {Keiser}, {Silbergleit}, {Holmes},
  {Kolodziejczak}, {Al-Meshari}, {Mester}, {Muhlfelder}, {Solomonik}, {Stahl},
  {Worden}, {Bencze}, {Buchman}, {Clarke}, {Al-Jadaan}, {Al-Jibreen}, {Li},
  {Lipa}, {Lockhart}, {Al-Suwaidan}, {Taber}, \& {Wang}}]{2011PhRvL.106v1101E}
{Everitt} C.~W.~F. {et~al.}, 2011, PhRvL, 106, 221101

\bibitem[{{Everitt} {et~al}\mbox{.}(2015){Everitt}, {Muhlfelder}, {DeBra},
  {Parkinson}, {Turneaure}, {Silbergleit}, {Acworth}, {Adams}, {Adler},
  {Bencze}, {Berberian}, {Bernier}, {Bower}, {Brumley}, {Buchman}, {Burns},
  {Clarke}, {Conklin}, {Eglington}, {Green}, {Gutt}, {Gwo}, {Hanuschak}, {He},
  {Heifetz}, {Hipkins}, {Holmes}, {Kahn}, {Keiser}, {Kozaczuk}, {Langenstein},
  {Li}, {Lipa}, {Lockhart}, {Luo}, {Mandel}, {Marcelja}, {Mester}, {Ndili},
  {Ohshima}, {Overduin}, {Salomon}, {Santiago}, {Shestople}, {Solomonik},
  {Stahl}, {Taber}, {Van Patten}, {Wang}, {Wade}, {Worden}, {Bartel}, {Herman},
  {Lebach}, {Ratner}, {Ransom}, {Shapiro}, {Small}, {Stroozas}, {Geveden},
  {Goebel}, {Horack}, {Kolodziejczak}, {Lyons}, {Olivier}, {Peters}, {Smith},
  {Till}, {Wooten}, {Reeve}, {Anderson}, {Bennett}, {Burns}, {Dougherty},
  {Dulgov}, {Frank}, {Huff}, {Katz}, {Kirschenbaum}, {Mason}, {Murray},
  {Parmley}, {Ratner}, {Reynolds}, {Rittmuller}, {Schweiger}, {Shehata},
  {Triebes}, {VandenBeukel}, {Vassar}, {Al-Saud}, {Al-Jadaan}, {Al-Jibreen},
  {Al-Meshari}, \& {Al-Suwaidan}}]{2015CQGra..32v4001E}
{Everitt} C.~W.~F. {et~al.}, 2015, CQGra, 32, 224001

\bibitem[{{Fokker}(1920)}]{1921KNAB...23..729F}
{Fokker} A.~D., 1920, KNAB, 29, 611

\bibitem[{{Genova} {et~al}\mbox{.}(2018){Genova}, {Mazarico}, {Goossens},
  {Lemoine}, {Neumann}, {Smith}, \& {Zuber}}]{2018NatCo...9..289G}
{Genova} A., {Mazarico} E., {Goossens} S., {Lemoine} F.~G., {Neumann} G.~A.,
  {Smith} D.~E., {Zuber} M.~T., 2018, NatCo, 9, 289

\bibitem[{{Giacobbe} {et~al}\mbox{.}(2020){Giacobbe}, {Benedetto}, {Damasso},
  {Sozzetti}, {Christille}, {Lattanzi}, {Calcidese}, {Carbognani}, {Barbato},
  {Pinamonti}, {Poggio}, {Lanza}, {Bernagozzi}, {Cenadelli}, {Lanteri}, \&
  {Bertolini}}]{2020MNRAS.491.5216G}
{Giacobbe} P. {et~al.}, 2020, \mnras, 491, 5216

\bibitem[{{Gould}, {Bahcall} \& {Flynn}(1996){Gould}, {Bahcall}, \&
  {Flynn}}]{1996ApJ...465..759G}
{Gould} A., {Bahcall} J.~N., {Flynn} C., 1996, \apj, 465, 759

\bibitem[{{G{\"u}nther} {et~al}\mbox{.}(2020){G{\"u}nther}, {Zhan}, {Seager},
  {Rimmer}, {Ranjan}, {Stassun}, {Oelkers}, {Daylan}, {Newton}, {Kristiansen},
  {Olah}, {Gillen}, {Rappaport}, {Ricker}, {Vanderspek}, {Latham}, {Winn},
  {Jenkins}, {Glidden}, {Fausnaugh}, {Levine}, {Dittmann}, {Quinn},
  {Krishnamurthy}, \& {Ting}}]{2020AJ....159...60G}
{G{\"u}nther} M.~N. {et~al.}, 2020, \aj, 159, 60

\bibitem[{{Haas} \& {Ross}(1975)}]{1975Ap&SS..32....3H}
{Haas} M.~R., {Ross} D.~K., 1975, Ap\&SS, 32, 3

\bibitem[{{Heath} {et~al}\mbox{.}(1999){Heath}, {Doyle}, {Joshi}, \&
  {Haberle}}]{1999OLEB...29..405H}
{Heath} M.~J., {Doyle} L.~R., {Joshi} M.~M., {Haberle} R.~M., 1999, OLEB, 29,
  405

\bibitem[{{Hofmann} \& {M{\"u}ller}(2018)}]{2018CQGra..35c5015H}
{Hofmann} F., {M{\"u}ller} J., 2018, CQGra, 35, 035015

\bibitem[{{Irwin} \& {Schulze-Makuch}(2020)}]{2020Univ....6..130I}
{Irwin} L.~N., {Schulze-Makuch} D., 2020, Univ, 6, 130

\bibitem[{{Kaltenegger}(2017)}]{2017ARA&A..55..433K}
{Kaltenegger} L., 2017, ARA\&A, 55, 433

\bibitem[{{Kaula}(1964)}]{1964RvGSP...2..661K}
{Kaula} W.~M., 1964, RvGSP, 2, 661

\bibitem[{{Kerr}(1987)}]{Kerr1987}
{Kerr} R.~A., 1987, Sci, 235, 973

\bibitem[{{Kilic}, {Raible} \& {Stocker}(2017){Kilic}, {Raible}, \&
  {Stocker}}]{2017ApJ...844..147K}
{Kilic} C., {Raible} C.~C., {Stocker} T.~F., 2017, ApJ, 844, 147

\bibitem[{{Kiraga} \& {Stepien}(2007)}]{2007AcA....57..149K}
{Kiraga} M., {Stepien} K., 2007, AcA, 57, 149

\bibitem[{{Kopal}(1959)}]{1959cbs..book.....K}
{Kopal} Z., 1959, {Close binary systems}. Chapman \& Hall, London

\bibitem[{{Kramer}(2012)}]{Kramer2012}
{Kramer} M., 2012, in The Twelfth Marcel Grossmann Meeting. Proceedings of the
  MG12 Meeting on General Relativity, {Damour} T., {Jantzen} R., {Ruffini} R.,
  eds., World Scientific, Singapore, pp. 241--260

\bibitem[{{Landin}, {Mendes} \& {Vaz}(2009){Landin}, {Mendes}, \&
  {Vaz}}]{2009A&A...494..209L}
{Landin} N.~R., {Mendes} L.~T.~S., {Vaz} L.~P.~R., 2009, \aap, 494, 209

\bibitem[{{Laskar}, {Joutel} \& {Robutel}(1993){Laskar}, {Joutel}, \&
  {Robutel}}]{1993Natur.361..615L}
{Laskar} J., {Joutel} F., {Robutel} P., 1993, Natur, 361, 615

\bibitem[{{Laskar} {et~al}\mbox{.}(2004){Laskar}, {Robutel}, {Joutel}, \& {et
  al.}}]{2004A&A...428..261L}
{Laskar} J., {Robutel} P., {Joutel} F., {et al.}, 2004, A\&A, 428, 261

\bibitem[{{Leconte}, {Lai} \& {Chabrier}(2011){Leconte}, {Lai}, \&
  {Chabrier}}]{2011A&A...528A..41L}
{Leconte} J., {Lai} D., {Chabrier} G., 2011, \aap, 528, A41

\bibitem[{{Lense} \& {Thirring}(1918)}]{1918PhyZ...19..156L}
{Lense} J., {Thirring} H., 1918, PhyZ, 19, 156

\bibitem[{{Li} \& {Batygin}(2014)}]{2014ApJ...790...69L}
{Li} G., {Batygin} K., 2014, ApJ, 790, 69

\bibitem[{{Lingam}, {Ginsburg} \& {Loeb}(2020){Lingam}, {Ginsburg}, \&
  {Loeb}}]{2020ApJ...888..102L}
{Lingam} M., {Ginsburg} I., {Loeb} A., 2020, \apj, 888, 102

\bibitem[{{Lingam} \& {Loeb}(2019)}]{2019RvMP...91b1002L}
{Lingam} M., {Loeb} A., 2019, RvMP, 91, 021002

\bibitem[{{Lingam} \& {Loeb}(2020)}]{2020IJAsB..19..379L}
{Lingam} M., {Loeb} A., 2020, IJAsB, 19, 379

\bibitem[{{Linsenmeier}, {Pascale} \& {Lucarini}(2015){Linsenmeier}, {Pascale},
  \& {Lucarini}}]{2015P&SS..105...43L}
{Linsenmeier} M., {Pascale} S., {Lucarini} V., 2015, P$\&$SS, 105, 43

\bibitem[{{Lissauer}(2012)}]{2012NewAR..56....1L}
{Lissauer} J.~J., 2012, NewAR, 56, 1

\bibitem[{{Lissauer}, {Barnes} \& {Chambers}(2012){Lissauer}, {Barnes}, \&
  {Chambers}}]{2012Icar..217...77L}
{Lissauer} J.~J., {Barnes} J.~W., {Chambers} J.~E., 2012, Icar, 217, 77

\bibitem[{{Mashhoon}(2001)}]{2001rfg..conf..121M}
{Mashhoon} B., 2001, in Reference Frames and Gravitomagnetism,
  {Pascual-S{\'a}nchez} J.~F., {Flor{\'\i}a} L., {San Miguel} A., {Vicente} F.,
  eds., World Scientific, Singapore, pp. 121--132

\bibitem[{{Maxted}(2018)}]{2018haex.bookE..18M}
{Maxted} P. F.~L., 2018, in Handbook of Exoplanets, {Deeg} H.~J., {Belmonte}
  J.~A., eds., p.~18

\bibitem[{{Mayor} \& {Queloz}(1995)}]{1995Natur.378..355M}
{Mayor} M., {Queloz} D., 1995, Natur, 378, 355

\bibitem[{{McKeegan}, {Kudryavtsev} \& {Schopf}(2007){McKeegan}, {Kudryavtsev},
  \& {Schopf}}]{2007Geo....35..591M}
{McKeegan} K.~D., {Kudryavtsev} A.~B., {Schopf} J.~W., 2007, Geo, 35, 591

\bibitem[{Meadows \& Barnes(2018)}]{Meadows2018}
Meadows V.~S., Barnes R.~K., 2018, in Handbook of Exoplanets, Deeg H.~J.,
  Belmonte J.~A., eds., Springer International Publishing, Cham, pp. 1--24

\bibitem[{{Mera}, {Chabrier} \& {Baraffe}(1996){Mera}, {Chabrier}, \&
  {Baraffe}}]{1996ApJ...459L..87M}
{Mera} D., {Chabrier} G., {Baraffe} I., 1996, \apjl, 459, L87

\bibitem[{{Milankovitch}(1941)}]{Milan1941}
{Milankovitch} M., 1941, Kanon der Erdbestrahlung und seine Anwendung auf das
  Eiszeitenproblem. Belgrad K\"{o}nigliche Serbische Akademie

\bibitem[{{Misner}, {Thorne} \& {Wheeler}(2017){Misner}, {Thorne}, \&
  {Wheeler}}]{2017grav.book.....M}
{Misner} C.~W., {Thorne} K.~S., {Wheeler} J.~A., 2017, {Gravitation}. Princeton
  University Press, Princeton

\bibitem[{{Mitrovica} \& {Forte}(1995)}]{1995GeoJI.121...21M}
{Mitrovica} J.~X., {Forte} A.~M., 1995, GeoJI, 121, 21

\bibitem[{{Mojzsis} {et~al}\mbox{.}(1996){Mojzsis}, {Arrhenius}, {McKeegan},
  {Harrison}, {Nutman}, \& {Friend}}]{1996Natur.384...55M}
{Mojzsis} S.~J., {Arrhenius} G., {McKeegan} K.~D., {Harrison} T.~M., {Nutman}
  A.~P., {Friend} C.~R.~L., 1996, Natur, 384, 55

\bibitem[{{Nakajima} {et~al}\mbox{.}(1995){Nakajima}, {Oppenheimer},
  {Kulkarni}, {Golimowski}, {Matthews}, \& {Durrance}}]{1995Natur.378..463N}
{Nakajima} T., {Oppenheimer} B.~R., {Kulkarni} S.~R., {Golimowski} D.~A.,
  {Matthews} K., {Durrance} S.~T., 1995, Natur, 378, 463

\bibitem[{{Newton} {et~al}\mbox{.}(2016){Newton}, {Irwin}, {Charbonneau},
  {Berta-Thompson}, {Dittmann}, \& {West}}]{2016ApJ...821...93N}
{Newton} E.~R., {Irwin} J., {Charbonneau} D., {Berta-Thompson} Z.~K.,
  {Dittmann} J.~A., {West} A.~A., 2016, \apj, 821, 93

\bibitem[{{Newton} {et~al}\mbox{.}(2018){Newton}, {Mondrik}, {Irwin},
  {Winters}, \& {Charbonneau}}]{2018AJ....156..217N}
{Newton} E.~R., {Mondrik} N., {Irwin} J., {Winters} J.~G., {Charbonneau} D.,
  2018, \aj, 156, 217

\bibitem[{{Nutman} {et~al}\mbox{.}(2016){Nutman}, {Bennett}, {Friend}, {van
  Kranendonk}, \& {Chivas}}]{2016Natur.537..535N}
{Nutman} A.~P., {Bennett} V.~C., {Friend} C. R.~L., {van Kranendonk} M.~J.,
  {Chivas} A.~R., 2016, Natur, 537, 535

\bibitem[{{Pais} {et~al}\mbox{.}(1999){Pais}, {Le Mou{\"e}l}, {Lambeck}, \&
  {Poirier}}]{1999E&PSL.174..155P}
{Pais} M.~A., {Le Mou{\"e}l} J.~L., {Lambeck} K., {Poirier} J.~P., 1999,
  E\&PSL, 174, 155

\bibitem[{{Perryman}(2018)}]{2018exha.book.....P}
{Perryman} M., 2018, {The Exoplanet Handbook. Second edition.} Cambridge Univ.
  Press, Cambridge

\bibitem[{{Pijpers}(1998)}]{1998MNRAS.297L..76P}
{Pijpers} F.~P., 1998, MNRAS, 297, L76

\bibitem[{{Poisson} \& {Will}(2014)}]{2014grav.book.....P}
{Poisson} E., {Will} C.~M., 2014, {Gravity}. Cambridge: Cambridge Univ. Press

\bibitem[{{Pugh}(1959)}]{Pugh59}
{Pugh} G., 1959, {Proposal for a Satellite Test of the Coriolis Prediction of
  General Relativity}. Research Memorandum~11, Weapons Systems Evaluation
  Group, The Pentagon, Washington D.C.

\bibitem[{{Quarles} {et~al}\mbox{.}(2019){Quarles}, {Barnes}, {Lissauer}, \&
  {Chambers}}]{2017arXiv171008052Q}
{Quarles} B., {Barnes} J.~W., {Lissauer} J.~J., {Chambers} J., 2019, AsBio, 20,
  73

\bibitem[{{Quarles}, {Li} \& {Lissauer}(2019){Quarles}, {Li}, \&
  {Lissauer}}]{2019arXiv191108431Q}
{Quarles} B., {Li} G., {Lissauer} J.~J., 2019, ApJ, 886, 56

\bibitem[{{Ragozzine} \& {Wolf}(2009)}]{2009ApJ...698.1778R}
{Ragozzine} D., {Wolf} A.~S., 2009, \apj, 698, 1778

\bibitem[{{Rappaport} {et~al}\mbox{.}(2014){Rappaport}, {Swift}, {Levine},
  {Joss}, {Sanchis-Ojeda}, {Barclay}, {Still}, {Handler}, {Ol{\'a}h},
  {Muirhead}, {Huber}, \& {Vida}}]{2014ApJ...788..114R}
{Rappaport} S. {et~al.}, 2014, \apj, 788, 114

\bibitem[{{Reiners} {et~al}\mbox{.}(2018){Reiners}, {Zechmeister}, {Caballero},
  {Ribas}, {Morales}, {Jeffers}, {Sch{\"o}fer}, {Tal-Or}, {Quirrenbach},
  {Amado}, {Kaminski}, {Seifert}, {Abril}, {Aceituno}, {Alonso-Floriano},
  {Ammler-von Eiff}, {Antona}, {Anglada-Escud{\'e}}, {Anwand-Heerwart},
  {Arroyo-Torres}, {Azzaro}, {Baroch}, {Barrado}, {Bauer}, {Becerril},
  {B{\'e}jar}, {Ben{\'\i}tez}, {Berdinas̃}, {Bergond}, {Bl{\"u}mcke},
  {Brinkm{\"o}ller}, {del Burgo}, {Cano}, {C{\'a}rdenas V{\'a}zquez}, {Casal},
  {Cifuentes}, {Claret}, {Colom{\'e}}, {Cort{\'e}s-Contreras}, {Czesla},
  {D{\'\i}ez-Alonso}, {Dreizler}, {Feiz}, {Fern{\'a}ndez}, {Ferro},
  {Fuhrmeister}, {Galad{\'\i}-Enr{\'\i}quez}, {Garcia-Piquer}, {Garc{\'\i}a
  Vargas}, {Gesa}, {G{\'o}mez Galera}, {Gonz{\'a}lez Hern{\'a}ndez},
  {Gonz{\'a}lez-Peinado}, {Gr{\"o}zinger}, {Grohnert}, {Gu{\`a}rdia},
  {Guenther}, {Guijarro}, {de Guindos}, {Guti{\'e}rrez-Soto}, {Hagen},
  {Hatzes}, {Hauschildt}, {Hedrosa}, {Helmling}, {Henning}, {Hermelo},
  {Hern{\'a}ndez Arab{\'\i}}, {Hern{\'a}ndez Casta{\~n}o}, {Hern{\'a}ndez
  Hernando}, {Herrero}, {Huber}, {Huke}, {Johnson}, {de Juan}, {Kim}, {Klein},
  {Kl{\"u}ter}, {Klutsch}, {K{\"u}rster}, {Lafarga}, {Lamert}, {Lamp{\'o}n},
  {Lara}, {Laun}, {Lemke}, {Lenzen}, {Launhardt}, {L{\'o}pez del Fresno},
  {L{\'o}pez-Gonz{\'a}lez}, {L{\'o}pez-Puertas}, {L{\'o}pez Salas},
  {L{\'o}pez-Santiago}, {Luque}, {Mag{\'a}n Madinabeitia}, {Mall}, {Mancini},
  {Mandel}, {Marfil}, {Mar{\'\i}n Molina}, {Maroto Fern{\'a}ndez},
  {Mart{\'\i}n}, {Mart{\'\i}n-Ruiz}, {Marvin}, {Mathar}, {Mirabet}, {Montes},
  {Moreno-Raya}, {Moya}, {Mundt}, {Nagel}, {Naranjo}, {Nortmann}, {Nowak},
  {Ofir}, {Oreiro}, {Pall{\'e}}, {Panduro}, {Pascual}, {Passegger}, {Pavlov},
  {Pedraz}, {P{\'e}rez-Calpena}, {P{\'e}rez Medialdea}, {Perger}, {Perryman},
  {Pluto}, {Rabaza}, {Ram{\'o}n}, {Rebolo}, {Redondo}, {Reffert}, {Reinhart},
  {Rhode}, {Rix}, {Rodler}, {Rodr{\'\i}guez}, {Rodr{\'\i}guez-L{\'o}pez},
  {Rodr{\'\i}guez Trinidad}, {Rohloff}, {Rosich}, {Sadegi},
  {S{\'a}nchez-Blanco}, {S{\'a}nchez Carrasco}, {S{\'a}nchez-L{\'o}pez},
  {Sanz-Forcada}, {Sarkis}, {Sarmiento}, {Sch{\"a}fer}, {Schmitt}, {Schiller},
  {Schweitzer}, {Solano}, {Stahl}, {Strachan}, {St{\"u}rmer}, {Su{\'a}rez},
  {Tabernero}, {Tala}, {Trifonov}, {Tulloch}, {Ulbrich}, {Veredas}, {Vico
  Linares}, {Vilardell}, {Wagner}, {Winkler}, {Wolthoff}, {Xu}, {Yan}, \&
  {Zapatero Osorio}}]{2018A&A...612A..49R}
{Reiners} A. {et~al.}, 2018, \aap, 612, A49

\bibitem[{{Rindler}(2001)}]{2001rsgc.book.....R}
{Rindler} W., 2001, {Relativity: special, general, and cosmological}. Oxford
  University Press, Oxford, UK

\bibitem[{{Sasaki} \& {Barnes}(2014)}]{2014IJAsB..13..324S}
{Sasaki} T., {Barnes} J.~W., 2014, IJAsB, 13, 324

\bibitem[{{Schiff}(1960)}]{Schiff60}
{Schiff} L., 1960, PhRvL, 4, 215

\bibitem[{{Schouten}(1918)}]{1918KNAB...27.214S}
{Schouten} W.~J.~A., 1918, KNAB, 27, 214

\bibitem[{{Schulze-Makuch} \& {Bains}(2018)}]{2018NatAs...2..432S}
{Schulze-Makuch} D., {Bains} W., 2018, NatAs, 2, 432

\bibitem[{{Schwieterman} {et~al}\mbox{.}(2018){Schwieterman}, {Kiang},
  {Parenteau}, {Harman}, {DasSarma}, {Fisher}, {Arney}, {Hartnett}, {Reinhard},
  {Olson}, {Meadows}, {Cockell}, {Walker}, {Grenfell}, {Hegde}, {Rugheimer},
  {Hu}, \& {Lyons}}]{2018AsBio..18..663S}
{Schwieterman} E.~W. {et~al.}, 2018, AsBio, 18, 663

\bibitem[{{Seager}(2011)}]{2010exop.book.....S}
{Seager} S., 2011, {Exoplanets}. University of Arizona Press, Tucson

\bibitem[{{Seager}(2013)}]{2013Sci...340..577S}
{Seager} S., 2013, Sci, 340, 577

\bibitem[{{Shahar} {et~al}\mbox{.}(2019){Shahar}, {Driscoll}, {Weinberger}, \&
  {Cody}}]{2019Sci...364..434S}
{Shahar} A., {Driscoll} P., {Weinberger} A., {Cody} G., 2019, Sci, 364, 434

\bibitem[{{Shan} \& {Li}(2018)}]{2018AJ....155..237S}
{Shan} Y., {Li} G., 2018, AJ, 155, 237

\bibitem[{{Shields}, {Ballard} \& {Johnson}(2016){Shields}, {Ballard}, \&
  {Johnson}}]{2016PhR...663....1S}
{Shields} A.~L., {Ballard} S., {Johnson} J.~A., 2016, PhR, 663, 1

\bibitem[{{Spiegel}, {Menou} \& {Scharf}(2009){Spiegel}, {Menou}, \&
  {Scharf}}]{2009ApJ...691..596S}
{Spiegel} D.~S., {Menou} K., {Scharf} C.~A., 2009, \apj, 691, 596

\bibitem[{{Sterne}(1939)}]{1939MNRAS..99..451S}
{Sterne} T.~E., 1939, \mnras, 99, 451

\bibitem[{{Su{\'a}rez Mascare{\~n}o}, {Rebolo} \& {Gonz{\'a}lez
  Hern{\'a}ndez}(2016){Su{\'a}rez Mascare{\~n}o}, {Rebolo}, \& {Gonz{\'a}lez
  Hern{\'a}ndez}}]{2016A&A...595A..12S}
{Su{\'a}rez Mascare{\~n}o} A., {Rebolo} R., {Gonz{\'a}lez Hern{\'a}ndez} J.~I.,
  2016, \aap, 595, A12

\bibitem[{{Tanner} {et~al}\mbox{.}(2012){Tanner}, {White}, {Bailey}, {Blake},
  {Blake}, {Cruz}, {Burgasser}, \& {Kraus}}]{2012ApJS..203...10T}
{Tanner} A., {White} R., {Bailey} J., {Blake} C., {Blake} G., {Cruz} K.,
  {Burgasser} A.~J., {Kraus} A., 2012, \apjs, 203, 10

\bibitem[{{Tarter} {et~al}\mbox{.}(2007){Tarter}, {Backus}, {Mancinelli},
  {Aurnou}, {Backman}, {Basri}, {Boss}, {Clarke}, {Deming}, {Doyle},
  {Feigelson}, {Freund}, {Grinspoon}, {Haberle}, {Hauck}, {Heath}, {Henry},
  {Hollingsworth}, {Joshi}, {Kilston}, {Liu}, {Meikle}, {Reid}, {Rothschild},
  {Scalo}, {Segura}, {Tang}, {Tiedje}, {Turnbull}, {Walkowicz}, {Weber}, \&
  {Young}}]{2007AsBio...7...30T}
{Tarter} J.~C. {et~al.}, 2007, AsBio, 7, 30

\bibitem[{{Teegarden} {et~al}\mbox{.}(2003){Teegarden}, {Pravdo}, {Hicks},
  {Lawrence}, {Shaklan}, {Covey}, {Fraser}, {Hawley}, {McGlynn}, \&
  {Reid}}]{2003ApJ...589L..51T}
{Teegarden} B.~J. {et~al.}, 2003, \apjl, 589, L51

\bibitem[{{Thorne}, {MacDonald} \& {Price}(1986){Thorne}, {MacDonald}, \&
  {Price}}]{Thorne86}
{Thorne} K.~S., {MacDonald} D.~A., {Price} R.~H., eds., 1986, {Black Holes: The
  Membrane Paradigm}. Yale University Press, Yale

\bibitem[{{Wang} {et~al}\mbox{.}(2016){Wang}, {Liu}, {Tian}, {Yang}, {Ding},
  {Zhou}, \& {Hu}}]{2016ApJ...823L..20W}
{Wang} Y., {Liu} Y., {Tian} F., {Yang} J., {Ding} F., {Zhou} L., {Hu} Y., 2016,
  \apjl, 823, L20

\bibitem[{{Ward}(1974)}]{1974JGR....79.3375W}
{Ward} W.~R., 1974, JGR, 79, 3375

\bibitem[{{West} {et~al}\mbox{.}(2015){West}, {Weisenburger}, {Irwin},
  {Berta-Thompson}, {Charbonneau}, {Dittmann}, \&
  {Pineda}}]{2015ApJ...812....3W}
{West} A.~A., {Weisenburger} K.~L., {Irwin} J., {Berta-Thompson} Z.~K.,
  {Charbonneau} D., {Dittmann} J., {Pineda} J.~S., 2015, \apj, 812, 3

\bibitem[{{Will}(2018)}]{2018tegp.book.....W}
{Will} C.~M., 2018, {Theory and Experiment in Gravitational Physics. Second
  edition}. Cabridge University Press, Cambridge

\bibitem[{{Williams} \& {Kasting}(1997)}]{1997Icar..129..254W}
{Williams} D.~M., {Kasting} J.~F., 1997, Icar, 129, 254

\bibitem[{{Williams} \& {Pollard}(2003)}]{2003IJAsB...2....1W}
{Williams} D.~M., {Pollard} D., 2003, IJAsB, 2, 1

\bibitem[{{Williams}(1975)}]{1975GeoM..112..441W}
{Williams} G.~E., 1975, GeoM, 112, 441

\bibitem[{{Williams} \& {Folkner}(2009)}]{2009IAU...261.0801W}
{Williams} J.~G., {Folkner} W.~M., 2009, in IAU Symposium \#261, American
  Astronomical Society, Vol. 261, p. 882

\bibitem[{{Williams}, {Newhall} \& {Dickey}(1996){Williams}, {Newhall}, \&
  {Dickey}}]{1996PhRvD..53.6730W}
{Williams} J.~G., {Newhall} X.~X., {Dickey} J.~O., 1996, PhRvD, 53, 6730

\bibitem[{{Wolszczan} \& {Frail}(1992)}]{1992Natur.355..145W}
{Wolszczan} A., {Frail} D.~A., 1992, Natur, 355, 145

\bibitem[{{Wright}(2018)}]{2018haex.bookE...4W}
{Wright} J.~T., 2018, in Handbook of Exoplanets, {Deeg} H.~J., {Belmonte}
  J.~A., eds., p.~4

\bibitem[{{Zechmeister} {et~al}\mbox{.}(2019){Zechmeister}, {Dreizler},
  {Ribas}, {Reiners}, {Caballero}, {Bauer}, {B{\'e}jar}, {Gonz{\'a}lez-Cuesta},
  {Herrero}, {Lalitha}, {L{\'o}pez-Gonz{\'a}lez}, {Luque}, {Morales},
  {Pall{\'e}}, {Rodr{\'\i}guez}, {Rodr{\'\i}guez L{\'o}pez}, {Tal-Or},
  {Anglada-Escud{\'e}}, {Quirrenbach}, {Amado}, {Abril}, {Aceituno},
  {Aceituno}, {Alonso-Floriano}, {Ammler-von Eiff}, {Antona Jim{\'e}nez},
  {Anwand-Heerwart}, {Arroyo-Torres}, {Azzaro}, {Baroch}, {Barrado},
  {Becerril}, {Ben{\'\i}tez}, {Berdi{\~n}as}, {Bergond}, {Bluhm},
  {Brinkm{\"o}ller}, {del Burgo}, {Calvo Ortega}, {Cano}, {Cardona
  Guill{\'e}n}, {Carro}, {C{\'a}rdenas V{\'a}zquez}, {Casal},
  {Casasayas-Barris}, {Casanova}, {Chaturvedi}, {Cifuentes}, {Claret},
  {Colom{\'e}}, {Cort{\'e}s-Contreras}, {Czesla}, {D{\'\i}ez-Alonso}, {Dorda},
  {Fern{\'a}ndez}, {Fern{\'a}ndez-Mart{\'\i}n}, {Fuhrmeister}, {Fukui},
  {Galad{\'\i}-Enr{\'\i}quez}, {Gallardo Cava}, {Garcia de la Fuente},
  {Garcia-Piquer}, {Garc{\'\i}a Vargas}, {Gesa}, {G{\'o}ngora Rueda},
  {Gonz{\'a}lez-{\'A}lvarez}, {Gonz{\'a}lez Hern{\'a}ndez},
  {Gonz{\'a}lez-Peinado}, {Gr{\"o}zinger}, {Gu{\`a}rdia}, {Guijarro}, {de
  Guindos}, {Hatzes}, {Hauschildt}, {Hedrosa}, {Helmling}, {Henning},
  {Hermelo}, {Hern{\'a}ndez Arabi}, {Hern{\'a}ndez Casta{\~n}o}, {Hern{\'a}ndez
  Otero}, {Hintz}, {Huke}, {Huber}, {Jeffers}, {Johnson}, {de Juan},
  {Kaminski}, {Kemmer}, {Kim}, {Klahr}, {Klein}, {Kl{\"u}ter}, {Klutsch},
  {Kossakowski}, {K{\"u}rster}, {Labarga}, {Lafarga}, {Llamas}, {Lamp{\'o}n},
  {Lara}, {Launhardt}, {L{\'a}zaro}, {Lodieu}, {L{\'o}pez del Fresno},
  {L{\'o}pez-Puertas}, {L{\'o}pez Salas}, {L{\'o}pez-Santiago}, {Mag{\'a}n
  Madinabeitia}, {Mall}, {Mancini}, {Mandel}, {Marfil}, {Mar{\'\i}n Molina},
  {Maroto Fern{\'a}ndez}, {Mart{\'\i}n}, {Mart{\'\i}n-Fern{\'a}ndez},
  {Mart{\'\i}n-Ruiz}, {Marvin}, {Mirabet}, {Monta{\~n}{\'e}s-Rodr{\'\i}guez},
  {Montes}, {Moreno-Raya}, {Nagel}, {Naranjo}, {Narita}, {Nortmann}, {Nowak},
  {Ofir}, {Oshagh}, {Panduro}, {Parviainen}, {Pascual}, {Passegger}, {Pavlov},
  {Pedraz}, {P{\'e}rez-Calpena}, {P{\'e}rez Medialdea}, {Perger}, {Perryman},
  {Rabaza}, {Ram{\'o}n Ballesta}, {Rebolo}, {Redondo}, {Reffert}, {Reinhardt},
  {Rhode}, {Rix}, {Rodler}, {Rodr{\'\i}guez Trinidad}, {Rosich}, {Sadegi},
  {S{\'a}nchez-Blanco}, {S{\'a}nchez Carrasco}, {S{\'a}nchez-L{\'o}pez},
  {Sanz-Forcada}, {Sarkis}, {Sarmiento}, {Sch{\"a}fer}, {Schmitt},
  {Sch{\"o}fer}, {Schweitzer}, {Seifert}, {Shulyak}, {Solano}, {Sota}, {Stahl},
  {Stock}, {Strachan}, {Stuber}, {St{\"u}rmer}, {Su{\'a}rez}, {Tabernero},
  {Tala Pinto}, {Trifonov}, {Veredas}, {Vico Linares}, {Vilardell}, {Wagner},
  {Wolthoff}, {Xu}, {Yan}, \& {Zapatero Osorio}}]{2019A&A...627A..49Z}
{Zechmeister} M. {et~al.}, 2019, \aap, 627, A49

\end{thebibliography}

\end{document}